\documentclass[12pt,preprint]{aastex}

\begin{document}
\newcommand{\W}{$W_0^{\lambda2796}$}

\title{The Environments of Ultra-strong Mg II
Absorbers\altaffilmark{1}}

\author{Daniel B. Nestor\altaffilmark{2}} \affil{Department of
Astronomy, University of Florida, Gainesville, FL, 32611}
\affil{Institute of Astronomy, Madingley Road, Cambridge CB3 0HA, UK}
\email{dbn@ast.cam.ac.uk} \and \author{David
A. Turnshek\altaffilmark{2}, Sandhya M. Rao\altaffilmark{2}, and Anna
M. Quider} \affil{Department of Physics \& Astronomy, University of
Pittsburgh,  Pittsburgh, PA 15260} \email{turnshek@pitt.edu;
rao@everest.phyast.pitt.edu; amq3@pitt.edu}

\altaffiltext{1}{Based on data obtained from the Sloan Digital Sky
Survey  (SDSS) and the WIYN$^3$ telescope.}

\altaffiltext{2}{Visiting Astronomer, Kitt Peak National Observatory,
National Optical Astronomy Observatory, which is operated by the
Association of Universities for Research in Astronomy, Inc. (AURA)
under cooperative agreement with the National Science Foundation.}

\altaffiltext{3}{The WIYN Observatory is a joint facility of the
University of Wisconsin-Madison, Indiana University, Yale University,
and the National Optical Astronomy Observatory.}

\begin{abstract}
We present $r^{\prime}$- or $i^{\prime}$-band WIYN\altaffilmark{3}
images of the fields of 15 Sloan Digital Sky Survey quasars that have
spectra exhibiting intervening \ion{Mg}{2} absorption-line systems
with rest equivalent widths 2.7 \AA\ $\le $\W $ \le 6.0$ \AA\ and
redshifts $0.42 < z_{abs} < 0.84$.  Such systems are rare and exhibit
projected absorption velocity spreads in excess of $\approx$ 300-650
km s$^{-1}$.  Approximately 60\% are expected to be damped Ly$\alpha$
systems.  In each of our fields we detect at least one galaxy that, if
at the absorption redshift, would have impact parameter $b \lesssim
40$ kpc and luminosity $L \gtrsim 0.3 L^*$.  We measure a significant
excess of galaxies at low-$b$ to the sightlines over a large range of
luminosity.  Many of the sightlines are found to pass either through
or close to the optically-luminous extent of a galaxy.  Considering
the very large velocity spreads seen in absorption, this suggests that
these absorbing regions are more kinematically complex than local
spirals such as the Milky Way.  Our data indicate that interactions
and galaxy pairs may be a contributing factor to the production of
such large velocity spreads.  Finally, we also find evidence that a
population of galaxies with luminosities in the range $4L^* \lesssim L
\lesssim 13 L^*$ may contribute to the presence of ultra-strong
\ion{Mg}{2} absorption.  Thus, some of the absorbing galaxies may
represent a population intermediate to the very luminous high-redshift
Lyman break galaxies and the fainter local starburst population.
\end{abstract}


\section{Introduction}
By providing a huge spectroscopic quasar database, the Sloan Digital
Sky Survey (SDSS) has spurred a renewed interest in large quasar
absorption-line surveys (e.g., Reichard et al.\ 2003; Nestor,
Turnshek, \& Rao 2005, hereafter NTR05; Prochaska, Herbert-Fort, \&
Wolfe 2005; Rao, Turnshek, \& Nestor 2006; McDonald et al.\ 2006; York
et al.\ 2006).  Such surveys not only improve the statistics of quasar
absorption lines (in some cases by several orders of magnitude), but
their large sizes also allow for the discovery of rare systems.  For
example, surveys for intervening \ion{Mg}{2} absorbers such as NTR05
and Nestor et al.\ (in preparation) have discovered significant
numbers of rare ``ultra-strong'' \ion{Mg}{2} absorption systems,
having $\lambda2796$ rest-frame equivalent widths 3 \AA\
$\lesssim$\W$\lesssim 6$ \AA.  Systems of this strength are 50 to
$10^{3.5}$ times more rare than those with \W$=0.3$ \AA, which is a
customary division between ``weak'' and ``strong'' \ion{Mg}{2}
absorbers.  Previous spectroscopic surveys were of insufficient size
to detect a very useful number of \ion{Mg}{2} absorbers at these
strengths.  Therefore, ultra-strong absorbers represent a virtually
unstudied phenomena.

The nature of the environments of less-strong \ion{Mg}{2} absorbers
have been the subject of many studies.  Churchill, Kacprzak, \&
Steidel (2005) have reviewed our understanding of systems with 0.1
\AA\ $ \lesssim $\W$ \lesssim 3$ \AA.  These absorbers are found to be
associated with the halos of galaxies that span a range of
morphological type, luminosity and impact parameter, but tend towards
galaxies with $0.1 \lesssim L/L^* \lesssim 3$ and impact parameters 15
kpc $\lesssim b \lesssim$ 80 kpc.  Recent results (e.g., Tripp \&
Bowen 2005) also suggest that the covering fraction for strong
absorption at these impact parameters is significantly less than unity
(perhaps $\approx 0.5$). Moreover, Rao et al.\ (2006) have shown that
a substantial fraction of systems with \W\ $\gtrsim 0.5$ \AA\ are
damped Ly$\alpha$ (DLA) absorbers, which have neutral hydrogen column
densities $N(HI) \ge 2\times10^{20}$ atoms cm$^{-2}$.  Rao et al.\
(2003 and references therein) have discussed imaging results on $z<1$
DLA galaxies, finding that low-luminosity and/or low surface
brightness dwarfs contribute substantially to this subset of strong
\ion{Mg}{2} absorbers.  At least three DLAs with particularly strong
(2.5 \AA\ $ \lesssim $\W$ \lesssim 3.0$ \AA) \ion{Mg}{2} absorption
have been imaged: the Q0827+243 ($z_{abs}=0.525$) absorber is
associated with an $\approx L^*$ spiral at $b=36$ kpc (Rao et al.\
2003; Steidel et al.\ 2002); the Q1137+3907 absorber ($z_{abs}=0.719$)
is associated with a sub-$L^*$ star-forming galaxy at $b=18$ kpc and
one or two possible sub-$L^*$ neighbors (Lacy et al.\ 2003); and the
Q1209+107 absorber ($z_{abs}=0.630$) has a galaxy in the field that is
$\approx 2L^*$ at $b=11$ kpc (Le Brun et al.\ 1997), assuming it is at
the absorber redshift.

For strong and at least partially saturated absorption lines, rest
equivalent widths (REWs) give an approximate indication of the
projected velocity spread, $\Delta V_{REW}$, of the absorbing gas,
with completely black saturated $\lambda2796$ lines providing a
lower-limit of $\Delta V_{REW} = 107.2 \times$ \W km s$^{-1}$.  Thus,
ultra-strong \ion{Mg}{2} absorbers exhibit strong/saturated absorption
over a (lower-limit) projected velocity spread of $\Delta V_{REW}
\approx 320-640$ km s$^{-1}$.  This is significantly larger than the
kinematic spread seen in Galactic \ion{Mg}{2} absorbers in quasar
spectra, which have a projected velocity extent along a sightline of
$125 \pm 35$ km s$^{-1}$ (Savage et al.\ 2000).

Though few individual ultra-strong absorbers have been studied in
detail, some of their properties are known.  Zibetti et al.\
(submitted to ApJ) demonstrate that stronger absorbers are on average
bluer and at smaller impact parameter.  M\'{e}nard et al.\ (in
preparation)  show that the amount of reddening of the background
quasar by the absorber is strongly correlated with REW, and Nestor et
al.\ (2003) and Turnshek et al.\ (2005) demonstrate that the gas-phase
metallicity and depletion of refractory elements in \ion{Mg}{2}
absorbers are strongly correlated with \W.  Thus, these systems are,
on average, relatively metal-rich and dusty compared to weaker
absorbers.  Indeed, the most metal-rich known DLA absorbers all have 2
\AA\ $\lesssim $\W$ \lesssim$ 3 \AA.  However, the results of Rao et
al.\ (2006) and NTR05 indicate that, while it is likely that a large
fraction of ultra-strong \ion{Mg}{2} absorption systems (perhaps
$\gtrsim 60\%$) have neutral hydrogen column densities above the DLA
threshold of N(\ion{H}{1}) $\ge 10^{20.3}$ atoms cm$^{-2}$, most
(perhaps $\approx 90\%$) DLAs have \W $ \lesssim$ 3 \AA.  It is
therefore of  interest to compare the range of galaxy types associated
with these  two overlapping classes of absorption system.

Aside from the clear empirical connection to DLA absorbers, other
authors have speculated on the nature of the strongest \ion{Mg}{2}
absorbers.  As a possible explanation for \ion{Mg}{2} systems with 0.9
\AA\ $\lesssim $\W$ \lesssim 1.3$ \AA\  that exhibit ``double-trough''
absorption profiles in higher resolution data, Churchill et al.\
(2000) investigated the probability of a line of sight passing through
galaxy pairs such as the Milky Way and either the LMC or SMC.  In this
regard, if the strongest absorbers do indeed trace interacting pairs,
their incidence and properties could have important impact on our
understanding of hierarchical formation scenarios. In addition, Bond
et al.\ (2001) investigated the possibility that galactic superwinds
give rise to sets of absorbers with 1.8 \AA\ $\lesssim $\W$ \lesssim
2.8$ \AA.  Confirmation of this connection would provide an important
tool for the study of the evolution of global star formation and
star-forming galaxies. Nestor, Turnshek, \& Rao (2006) discuss
evidence in support of both superwinds and interactions as possible
mechanisms causing ultra-strong \ion{Mg}{2} absorbers.  Both of these
connections would also have potentially important impact on the
understanding of environmental effects such as the distribution  of
metals into the IGM (e.g., through galactic fountains).

While there are many examples of the association between less-strong
\ion{Mg}{2} absorption and galaxies with small impact parameters, the
sightlines do not in general pass through the optically luminous
extent of the galaxies.  The simplest possible explanation for
ultra-strong \ion{Mg}{2} absorbers may therefore be that these systems
select sightlines passing through the inner regions of the disks as
well as the halos of massive galaxies.  However, even if this is
indeed the case, the processes responsible for the very large
kinematic spreads would still require explanation.

Thus, depending on the nature of ultra-strong \ion{Mg}{2} absorbers,
they may have important implications for our understanding of DLA
galaxies, massive galaxies, galaxy clustering, major and minor
mergers, the evolution of starburst galaxies, and metals in the IGM.
It is clearly of interest to determine the physical processes
responsible for these rare and extreme systems with strong low-ion gas
absorption over such a large interval of rest-frame velocity.
Consequently, in order to constrain the nature of ultra-strong
\ion{Mg}{2} absorbers, we have initiated a program to study the
associated galaxies and environments through optical imaging. Targeted
follow-up spectroscopy of individual galaxies in the fields should
follow. Here we present WIYN $r^{\prime}$ or $i^{\prime}$ images of
the fields towards 15 quasars exhibiting \ion{Mg}{2} absorbers with
2.69 \AA\ $< $\W$ \lesssim 5.97$ \AA\ and $0.42 \le z_{abs} \le 0.84$,
and we discuss the properties of the sample of detected galaxies.  We
describe our observations in \S~\ref{section:obser}, the results in
\S~\ref{section:results}, discuss the implications of this work in
\S~\ref{section:discussion}, and present our conclusions in
\S~\ref{section:conclusions}.

\section{Observations} 
\label{section:obser}
We conducted observations at the WIYN telescope on Kitt Peak, Arizona,
using the WIYN Tip-Tilt Module and the SDSS $r^{\prime}$ and
$i^{\prime}$ filters. We imaged quasar fields containing strong and
ultra-strong \ion{Mg}{2} absorbers detected in SDSS quasar spectra.
For the eight absorbers with $0.42 < z_{abs} < 0.6$ we used the
$r^{\prime}$ filter, and for the seven absorbers with $0.6 < z_{abs} <
0.84$ we used the $i^{\prime}$ filter to ensure that the 4000 \AA\
break was not redshifted into the filter bandpass. We were awarded two
nights each in February of 2005 and January of 2006, although one and
a half nights were lost to weather, and one night had rather poor
seeing.  The final night, however, had fair to good seeing throughout.
In all, we were able to obtain usable images of 15 fields, though as
indicated below the seeing quality was quite heterogeneous.  We
obtained at least three images of each field in order to facilitate
removal of cosmic rays, and total exposure times were between 60 and
100 minutes per field. The quasar point spread function (PSF) was well
behaved in most of our images, and bright, unsaturated stars allowed
for the modelling and removal of the quasar PSF in all but two cases.

\section{Results}
\label{section:results}
We used the software package SExtractor (Bertin \& Arnouts, 1996) to
detect, deblend, and measure sources in each of our fields.  All
regions having at least 15 contiguous pixels with fluxes 1.5$\sigma$
above the background were extracted.  We calibrated our photometry
using SDSS imaging results on available bright objects in each field.
In Table~\ref{table:photometry} we list all resolved objects detected
within an impact parameter of $b \le 200$ kpc of the quasar sightlines
under the assumption that the objects are at the absorption redshift.
Uncertainties in the magnitudes were generally $\sigma_m < 0.1$ mag,
resulting in better than 10\% photometric accuracy.  All
cosmology-dependent quantities are computed assuming $\Omega_M = 0.3$,
$\Omega_\Lambda = 0.7$, and $h=0.7$.  Absolute magnitudes (and
luminosities) for our detections were determined using three different
$k$-corrections\footnote{Computed using the IRAF package COSMOPACK,
developed by Balcells \& Cristobal,
\url{http://www.iac.es/galeria/balcells/publ\_mbc.html}.} for galaxy
types Sc, Sa, and E.  Galaxy impact parameter, $b$, and sizes are also
given in Table~\ref{table:photometry}.  The apparent projected optical
extents (i.e., the luminous ``sizes'' of galaxies) are listed as the
major-axis $\times$ minor-axis of the SExtractor-fitted ellipse, which
corresponds to the source's approximate isophotal limit.  We note that
the ellipses drawn on each image in Figures 2-16 are the photometric
integration limits and correspond to twice the isophotal limit.

For ease of discussion, as well as to facilitate comparison of
absolute magnitudes independent of filter, we also list luminosities
relative to those for the ``characteristic'' absolute magnitudes,
$M^*$, from Schechter function fits to the galaxy luminosity function
(LF).  Whereas the actual calculated value of $M^*$ (or $L^*$) is
correlated with the calculated faint-end slope of the LF, and varies
from study to study (see table 1 of Brown et al., 2001, for example),
we stress that the the physically meaningful quantities are the
relative number-densities of galaxies as a function of luminosity,
which do not depend on the actual value of $M^*$.  For our purposes,
we adopt values for $M^*$ determined from the SDSS for the relatively
local universe (i.e., $0.02 < z < 0.22$) from Blanton et al., 2003.
In our cosmology, these correspond to $M^*_r = -21.21$ and $M^*_i =
-21.59$.  We note that these values are for SDSS filters $k$-corrected
to $z=0.1$, making them perhaps non-ideal for direct comparison.
However, these values have the advantages of being calculated from the
same data set to which we have calibrated our photometry and, more
importantly, have very similar number-densities (i.e., $\phi(M^*_r)
\simeq \phi(M^*_i)$) which allows for meaningful comparisons between
our two filters.  Furthermore, the value for $M^*_r$ matches that used
in at least some previous work (e.g., Rao et al., 2003).  Unless
otherwise stated, $L^*$ corresponds to these values throughout this
work.  For comparison, in the extremely local ($10 h^{-1}$ Mpc $< d <
150 h^{-1} $Mpc) universe, Blanton et al., 2005 find $M^*_r$ and
$M^*_i$ to be approximately 36\% and 47\% fainter than the above
values, respectively.

\subsection{Discussion of Individual Fields}
\begin{figure}
\epsscale{1.0}
\plotone{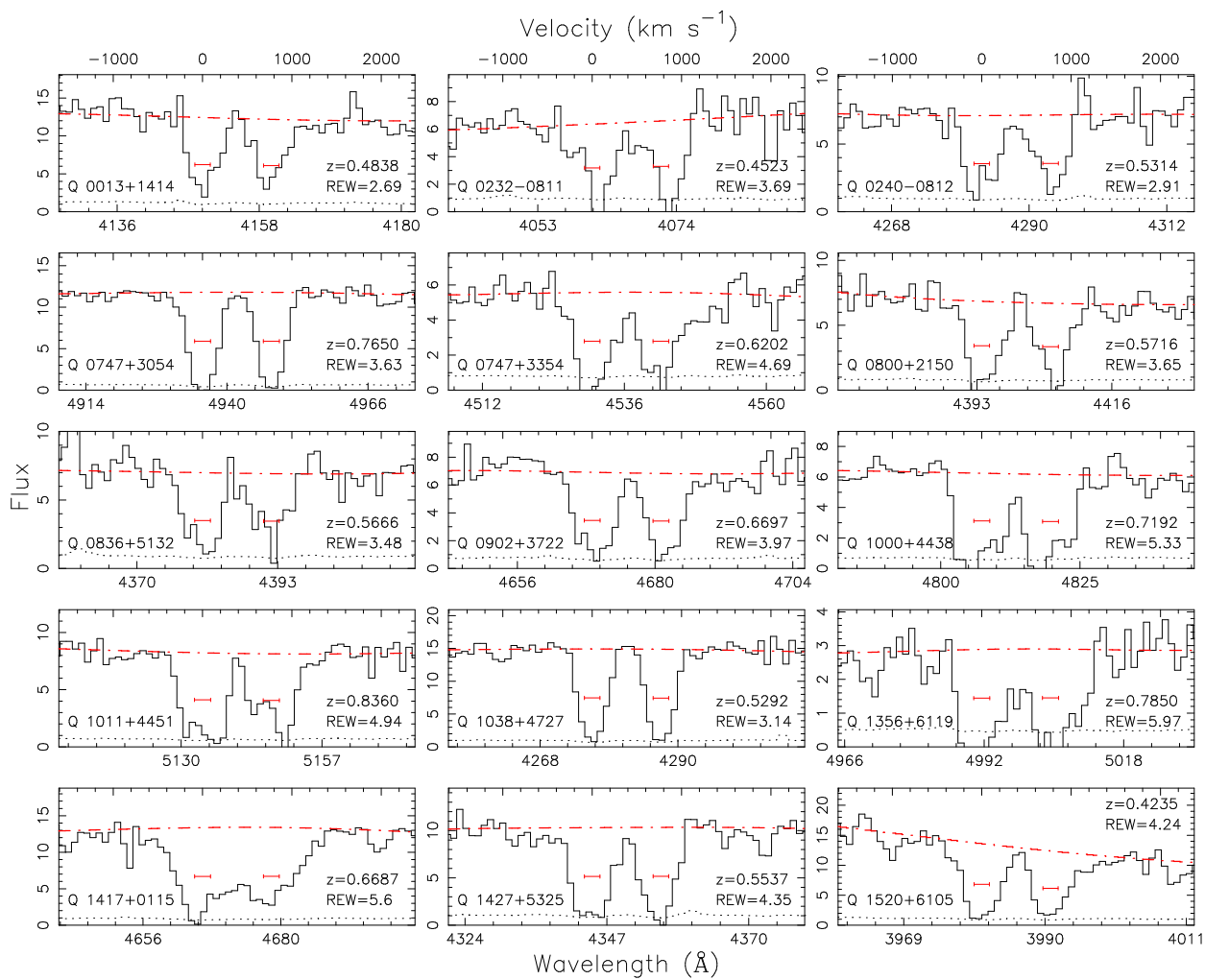}
\caption{Absorption region of the SDSS spectra containing ultra-strong
\ion{Mg}{2} absorption systems.  Horizontal bars indicate the
resolution FWHM.  The top x-axes show rest-frame velocity relative to 
$\lambda2796$.}
\label{specs}
\end{figure}

\begin{figure}
\plotone{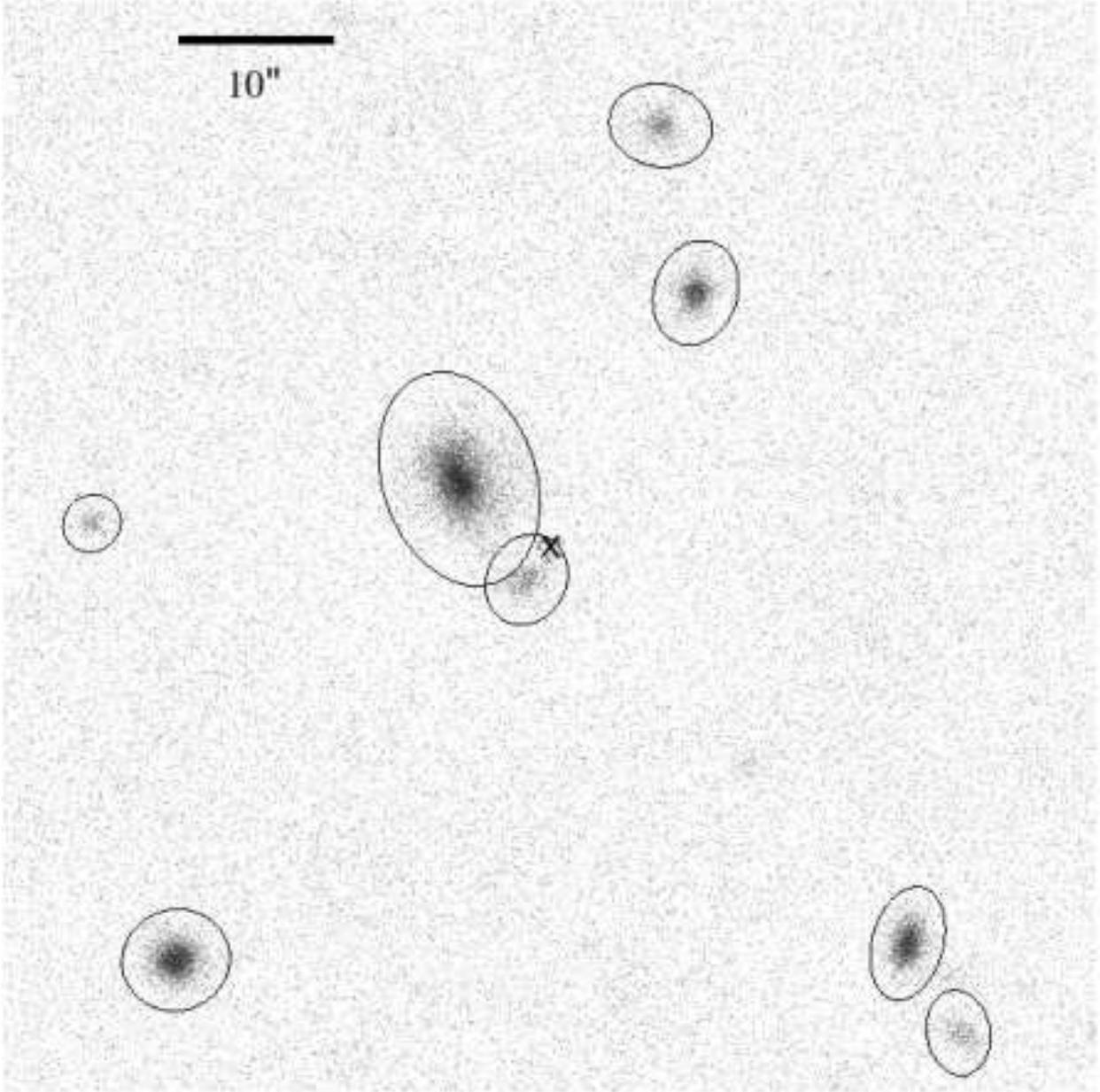}
\caption{WIYN $r^\prime$ image of the  Q0013+1414 field.  The
absorption has $z_{abs}=0.4838$ and \W$=2.69$ \AA.  At the absorption
redshift, $10^{\arcsec}$ corresponds to 60.0 kpc.  In the following figures,
ellipses represent photometric integration limits (twice the isophotal limit, see
\S~\ref{section:results}) and unless otherwise
noted the quasar PSF has been subtracted and
it's location marked with the symbol ``x''.}
\label{image:q0013}
\end{figure}

\begin{figure}
\plotone{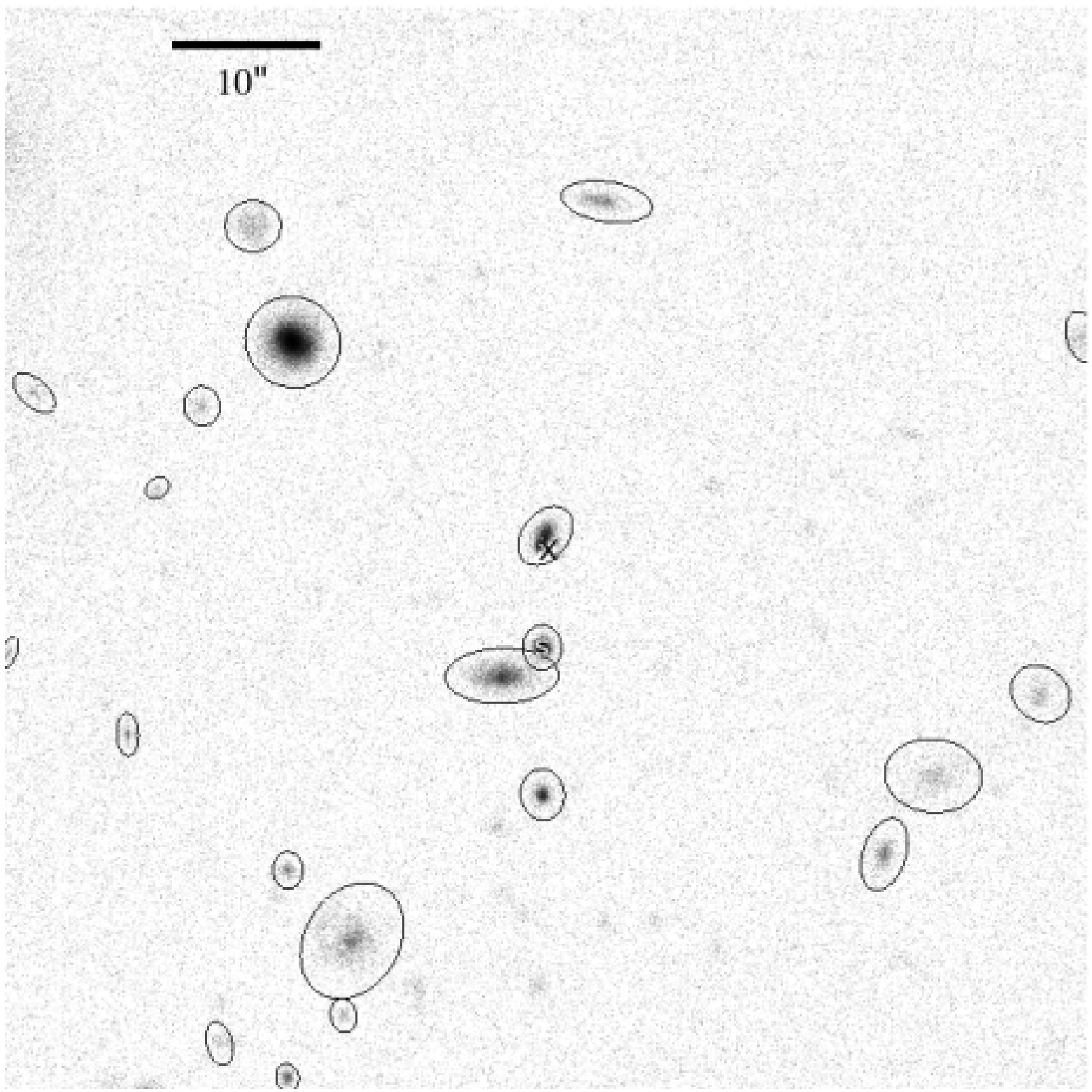}
\caption{WIYN $r^\prime$ image of the  Q0232-0811 field.  The absorption has
$z_{abs}=0.4523$ and \W$=3.69$ \AA.  At the absorption
redshift, $10^{\arcsec}$ corresponds to 57.8 kpc.}
\label{image:q0232}
\end{figure}

\begin{figure}
\plotone{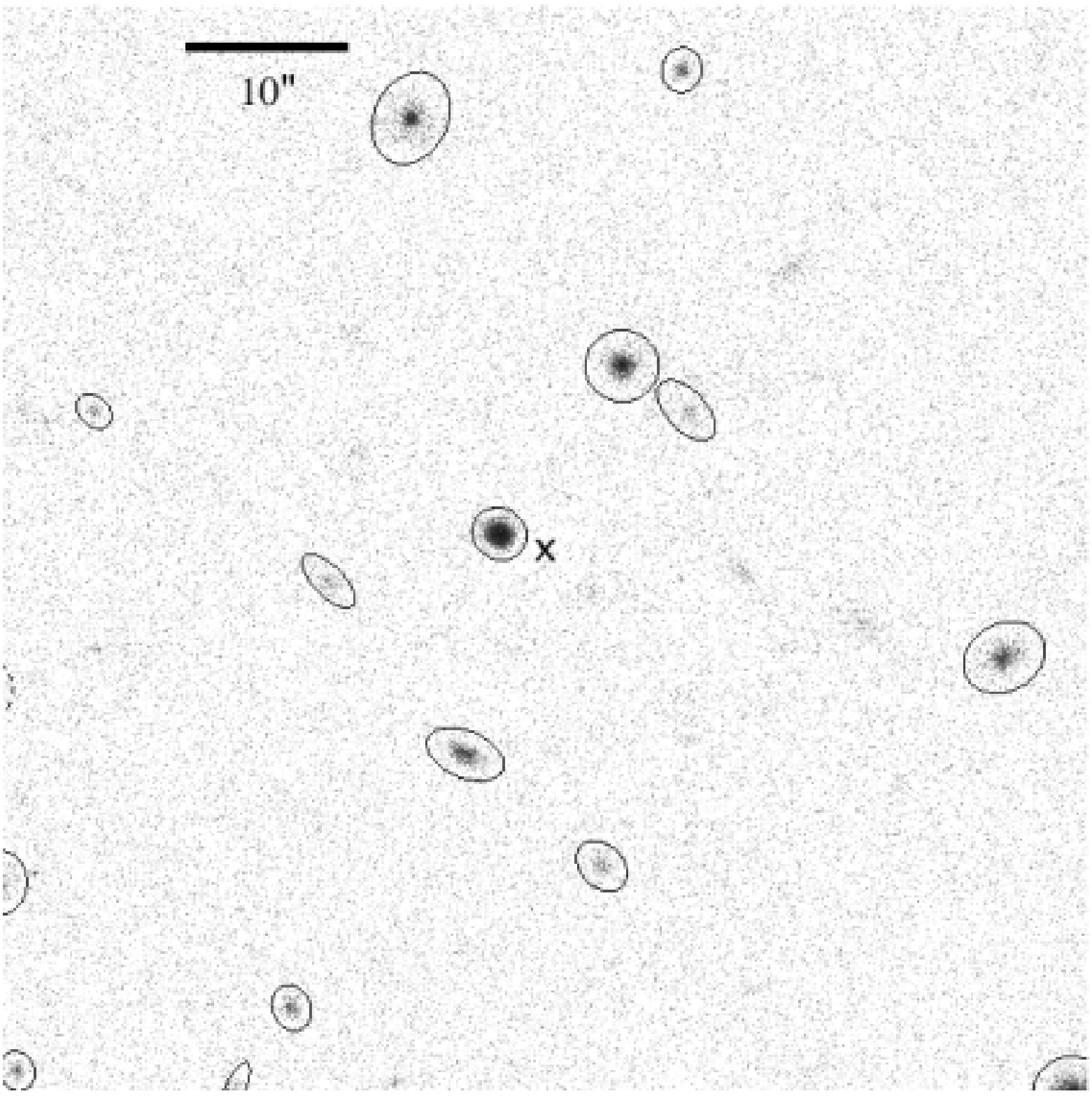}
\caption{WIYN $r^\prime$ image of the Q0240-0812 field.  The
absorption has $z_{abs}=0.5313$ and \W$=2.91$ \AA.  At the absorption
redshift, $10^{\arcsec}$ corresponds to 63.0 kpc.}
\label{image:q0240}
\end{figure}

The \ion{Mg}{2} absorption region of each spectrum is shown in
Figure~\ref{specs}.  Note that the SDSS spectra reach a minimum
\ion{Mg}{2} redshift of $z_{abs} = 0.36$.  Thus, there is no
absorption information available below this redshift.  Above this
value, the signal to noise ratios of the spectra were such that our
formal \W\ detection threshold was typically between 0.4 \AA\ and 1
\AA, though for $0.36 \le z \lesssim 0.5$ this value typically rises
to between 0.6\AA\ and 2\AA\ due to increased noise at the blue end of
the spectral coverage.  Figures~\ref{image:q0013} through
~\ref{image:q1520} show our WIYN images. All are centered on the
quasar sightlines and oriented such that north is to the top and east
is to the left. They are sized so that they span 420 kpc $\times$ 420
kpc at the absorption redshift.  The location of each sightline is
marked with a ``x'' and unresolved objects classified as stars are
labelled with an ``s''.  In all but two cases (Q0800+2150 and
Q0902+3722) the Point Spread Function (PSF) was determined well enough
from stars in the field that it was subtracted at the quasar position
(hence the quasar appears to disappear in 13 of the 15 fields).
Magnitudes and colors from the SDSS, when quoted, use the so-called
SDSS ``model'' magnitudes\footnote{See
\url{http://cas.sdss.org/dr5/en/help/docs/glossary.asp}.}.  For
convenience, quasar names are specified as coordinate designations
relative to equinox 2000, but the formal SDSS quasar names are also
listed in the text.

In our discussion of each field we emphasise what we consider to be
the most likely interpretation, keeping in mind that ultra-strong
\ion{Mg}{2} absorbers are rare and may select correspondingly rare
galaxy types and/or environments. Due to our lack of confirmed
redshifts for almost all of the galaxies in these fields, all of these
interpretations should be considered preliminary. However, in many
cases there is clearly good circumstantial evidence for our proposed
interpretation.

\subsubsection{Q0013+1414}
The spectrum of Q0013+1414 (SDSS J001335.75+141424.1, $z_{em}=1.541$)
contains a strong (\W$=2.69 \pm 0.14$ \AA) \ion{Mg}{2} absorber at
$z_{abs}=0.4838$.  No other low redshift ($z \lesssim 1$) \ion{Mg}{2}
absorber was detected in the spectrum.  Our WIYN $r^\prime$-band image
of the field is shown in Figure~\ref{image:q0013}.  The seeing was
poor during the observation (FWHM $\approx 1.3^{\arcsec}$) and we
reached a limiting surface brightness of $r^{\prime} = 24.18$ mag
arcsec$^{-2}$.  The galaxy 7$^{\arcsec}$ to the northeast of the
sightline has $r^{\prime} = 19.68$.  At the absorption redshift our
photometry corresponds to $L_r =$ 4.4/5.6/6.4 $L_r^*$ (for
Sc/Sa/E-type $k$-corrections), an impact parameter of $b=44$ kpc, and
a  projected optical extent of $\approx 50\times35$ kpc.  It is also
detected in the SDSS images with $r^{\prime} \approx 19.5$ and
$g^{\prime}-r^{\prime}\approx0.6$.  If at $z=z_{abs}$, its blue colors
would be indicative of a starbursting galaxy.  We were able to obtain
a very good quasar PSF subtraction and resolved light blended with the
quasar PSF was also detected.  We measured this residual light to have
$r^{\prime} \approx 22.0$, which corresponds to 0.5/0.7/0.8 $L_r^*$
and $b= 16$ kpc.  The most likely interpretation is that the low
impact parameter fainter source is associated with the absorption and
the brighter galaxy is in the foreground.  However, it is also
possible that the bright galaxy is at $z=z_{abs}$ with, perhaps, the
fainter galaxy as a satellite.


\subsubsection{Q0232$-$0811}
The spectrum of Q0232$-$0811 (SDSS J023234.06$-$081126.5,
$z_{em}=1.656$) contains an ultra-strong (\W=$3.69 \pm 0.28$ \AA)
\ion{Mg}{2} absorber at $z_{abs}=0.4523$.  No other low redshift
\ion{Mg}{2} absorption system was detected.  The profiles of both
lines of the \ion{Mg}{2} doublet, $\lambda2796$ and $\lambda2803$,
appear to exhibit wings $\approx 340$ km s$^{-1}$ blueward of the main
absorption.  Figure~\ref{image:q0232} shows an $r^\prime$-band image
of the field  obtained under fair seeing conditions (FWHM $=
0.69^{\arcsec}$) with a limiting surface brightness of $r^{\prime} =
24.59$ mag arcsec$^{-2}$.  We were able to remove the quasar PSF
fairly well with only moderate residuals.  This revealed a source that
had been blended with the quasar PSF, having $r^{\prime} \approx 22.1$
and $b \approx 0.8^{\arcsec}$, which corresponds to $L_r
\approx$0.4/0.5/0.6$L_r^*$ and $b \approx 5$ kpc.  The object
7$^{\arcsec}$ to the south of the sightline is unresolved and is
likely a star.  The galaxy 9$^{\arcsec}$ to the south has $r^{\prime}
= 21.4$,  which corresponds to $b=52$ kpc and $L_r
=$0.8/1.0/1.1$L_r^*$.  It is also detected in the SDSS with
$r^{\prime}\approx21.3$ and $g^{\prime}-r^{\prime} \approx 1.4$, which
is consistent with the colors of a local late-type galaxy shifted to
$z=z_{abs}$.  The presence of a source essentially covering the
sightline indicates that it is likely responsible for the ultra-strong
absorption.


\subsubsection{Q0240$-$0812}
The spectrum of Q0240$-$0812 (SDSS J024008.21$-$081223.4,
$z_{em}=2.231$) contains a strong (\W$=2.91 \pm 0.26$ \AA) \ion{Mg}{2}
absorber at $z_{abs}=0.5314$.  No other low redshift \ion{Mg}{2}
absorber was detected.  We obtained three WIYN $r^\prime$-band images.
However, only one had decent seeing (FWHM $= 0.69^{\arcsec}$), while
the others were obtained in poorer conditions (FWHM $\approx
1.3^{\arcsec}$).  The combined images reach a limiting surface
brightness of $r^{\prime} = 24.50$ mag arcsec$^{-2}$ to within $b
\simeq11.7$ kpc.  In Figure~\ref{image:q0240} we show just the best
image, with the quasar PSF removed.  This image reaches  $r^{\prime} =
23.71$ mag arcsec$^{-2}$.  The galaxy 3$^{\arcsec}$ to the east has
$r^{\prime} = 20.96$, which corresponds to $b=18$ kpc and $L_r
=$1.7/2.3/2.8$L_r^*$. It is also detected in the SDSS images with
$r^{\prime} \approx 21.6$ and $g^{\prime}-r^{\prime} \approx 0.7$,
which is rather blue for being at the absorption redshift, though we
note that the SDSS magnitudes might be less accurate due to the
brighter limiting surface brightness and blending with the quasar PSF.
We consider it highly likely that it is associated with the strong
absorption.

%

\subsubsection{Q0747+3054}
\begin{figure}
\plotone{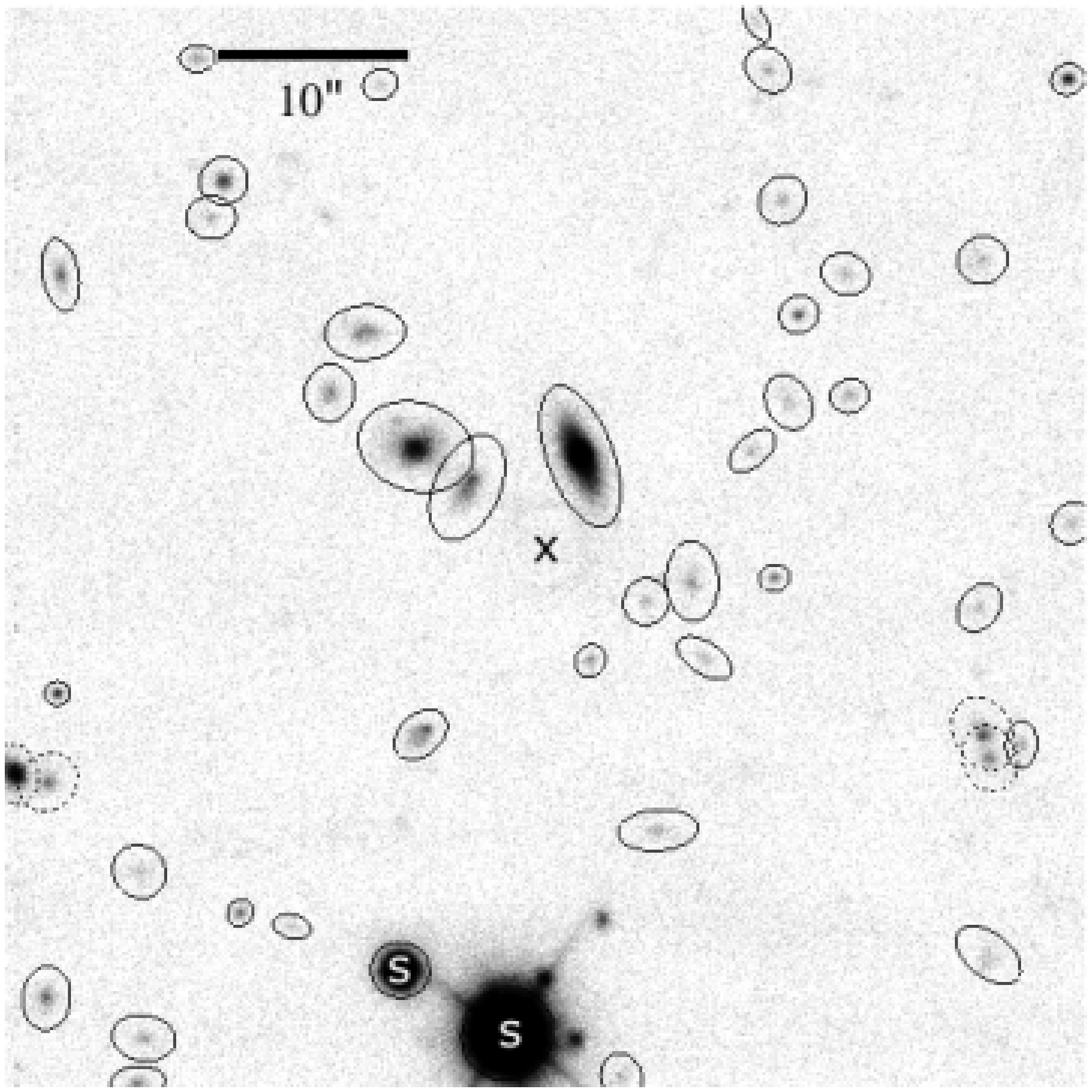}
\caption{WIYN $i^\prime$ image of the Q0747+3054 field.  The absorption
has $z_{abs}=0.7650$ and \W$=3.63$ \AA.  At the absorption
redshift, $10^{\arcsec}$ corresponds to 73.9 kpc.}
\label{image:q0747}
\end{figure}

\begin{figure}
\plotone{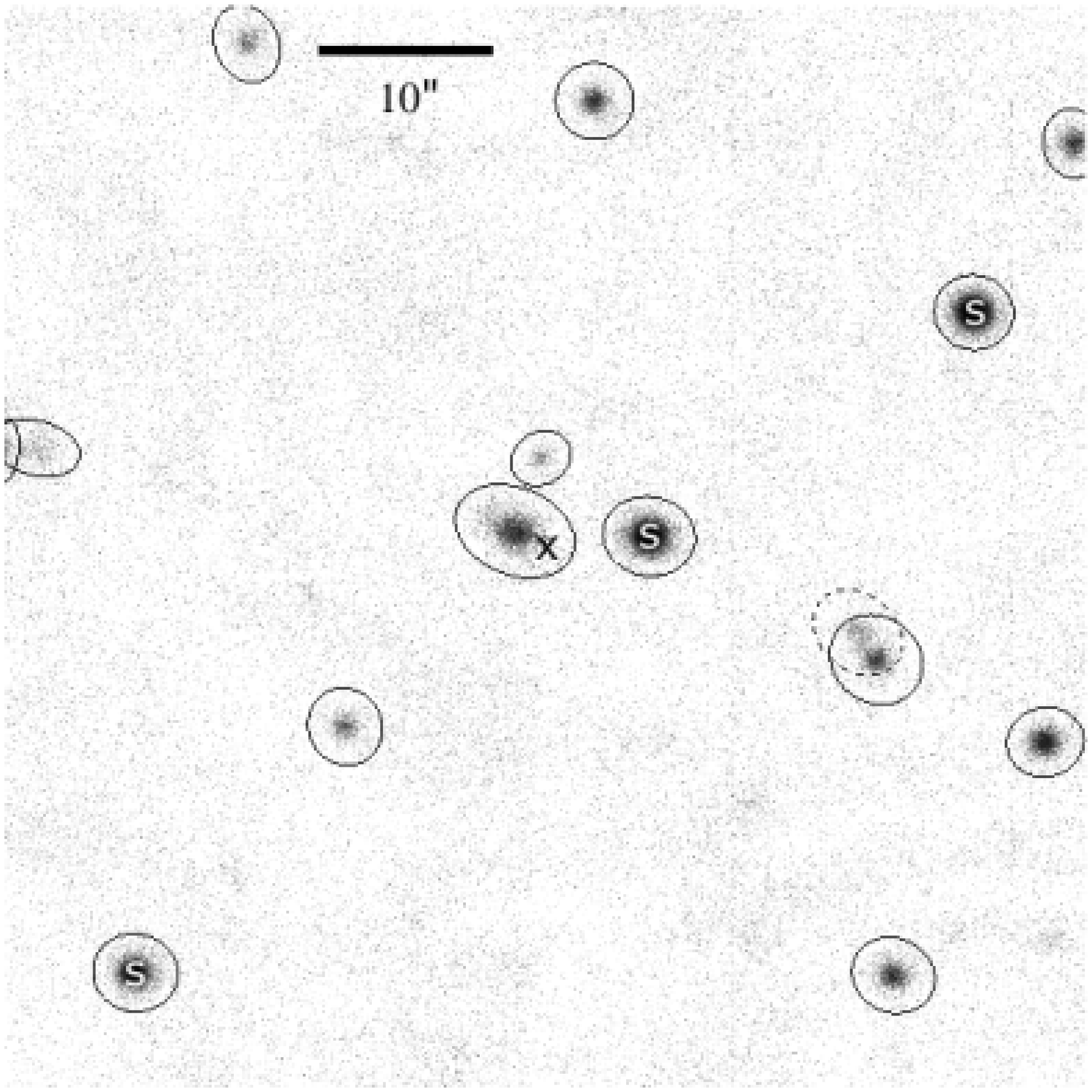}
\caption{WIYN $i^\prime$ image of the  Q0747+3354 field.  The
absorption has $z_{abs}=0.6202$ and \W$=4.69$ \AA.  At the absorption
redshift, $10^{\arcsec}$ corresponds to 67.9 kpc.}
\label{image:q0748}
\end{figure}

\begin{figure}
\plotone{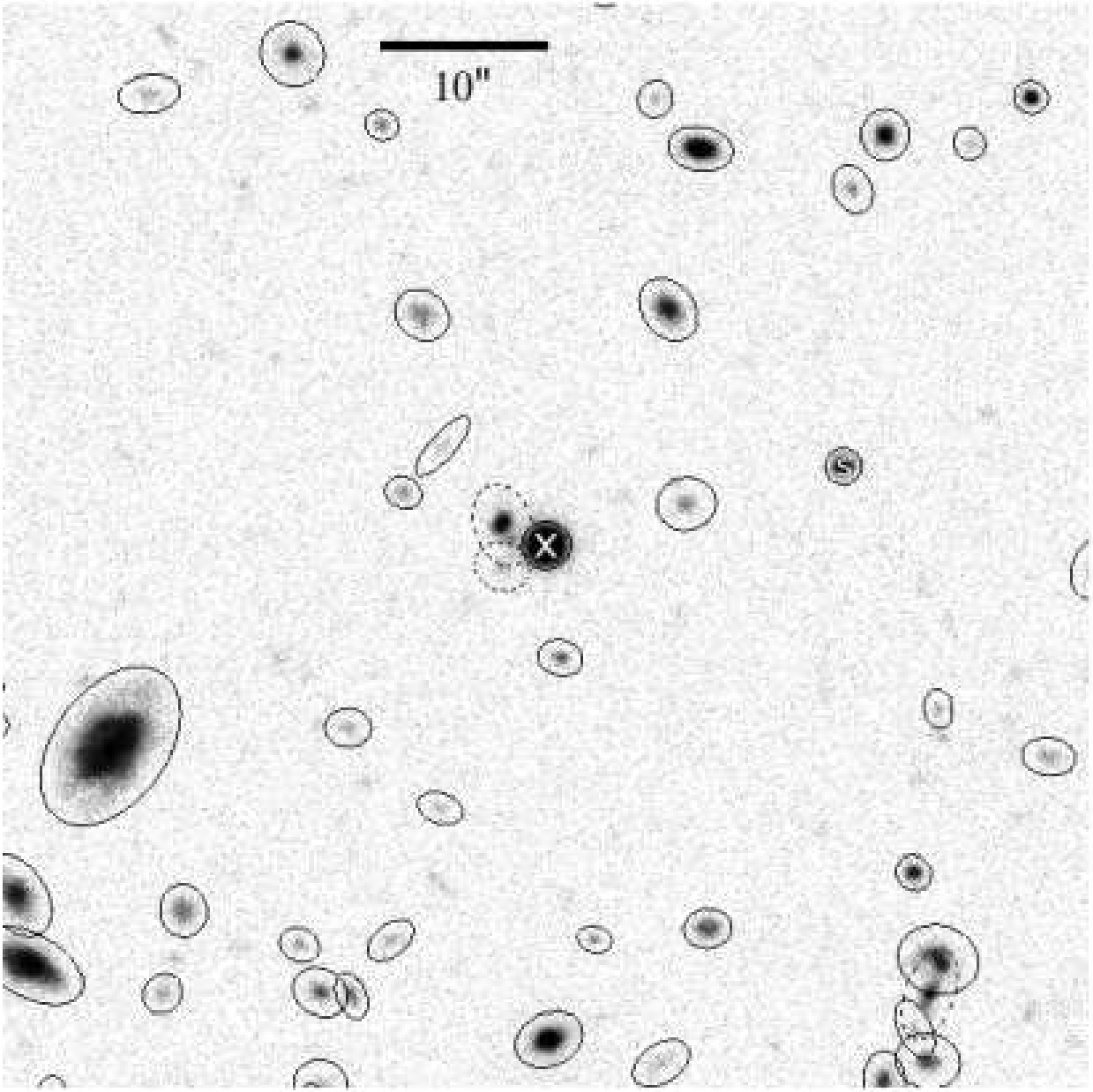}
\caption{WIYN $r^\prime$ image of the  Q0800+2150 field.  The absorption
has $z_{abs}=0.5716$ and \W$=3.65$ \AA.  At the absorption
redshift, $10^{\arcsec}$ corresponds to 65.3 kpc.  The quasar PSF has
not been subtracted in this image.}
\label{image:q0800}
\end{figure}

\begin{figure}
\plotone{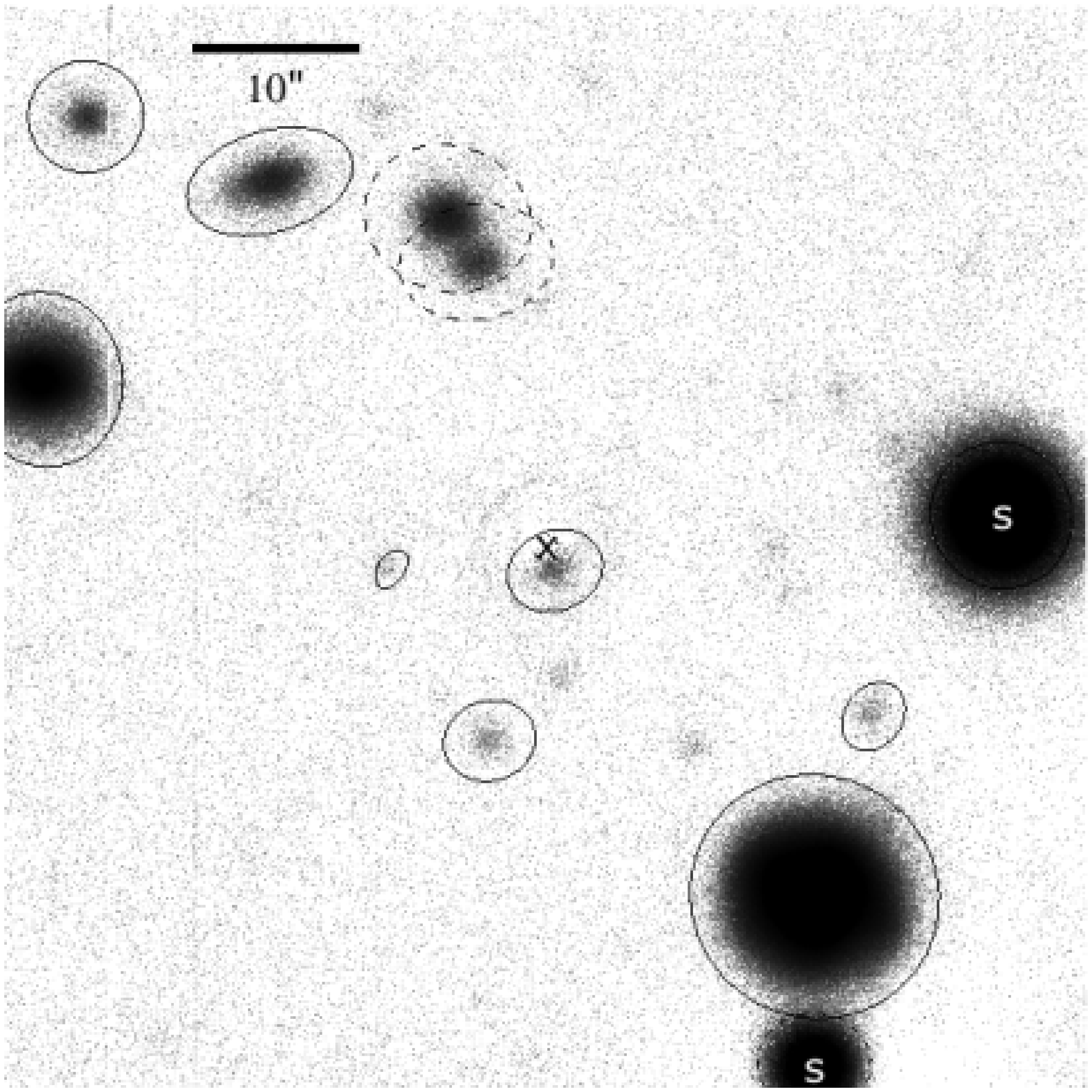}
\caption{WIYN $r^\prime$ image of the  Q0836+5132 field.  The
absorption has $z_{abs}=0.5666$ and \W$=3.48$ \AA.  At the absorption
redshift, $10^{\arcsec}$ corresponds to 65.1 kpc.}
\label{image:q0836}
\end{figure}

\begin{figure}
\plotone{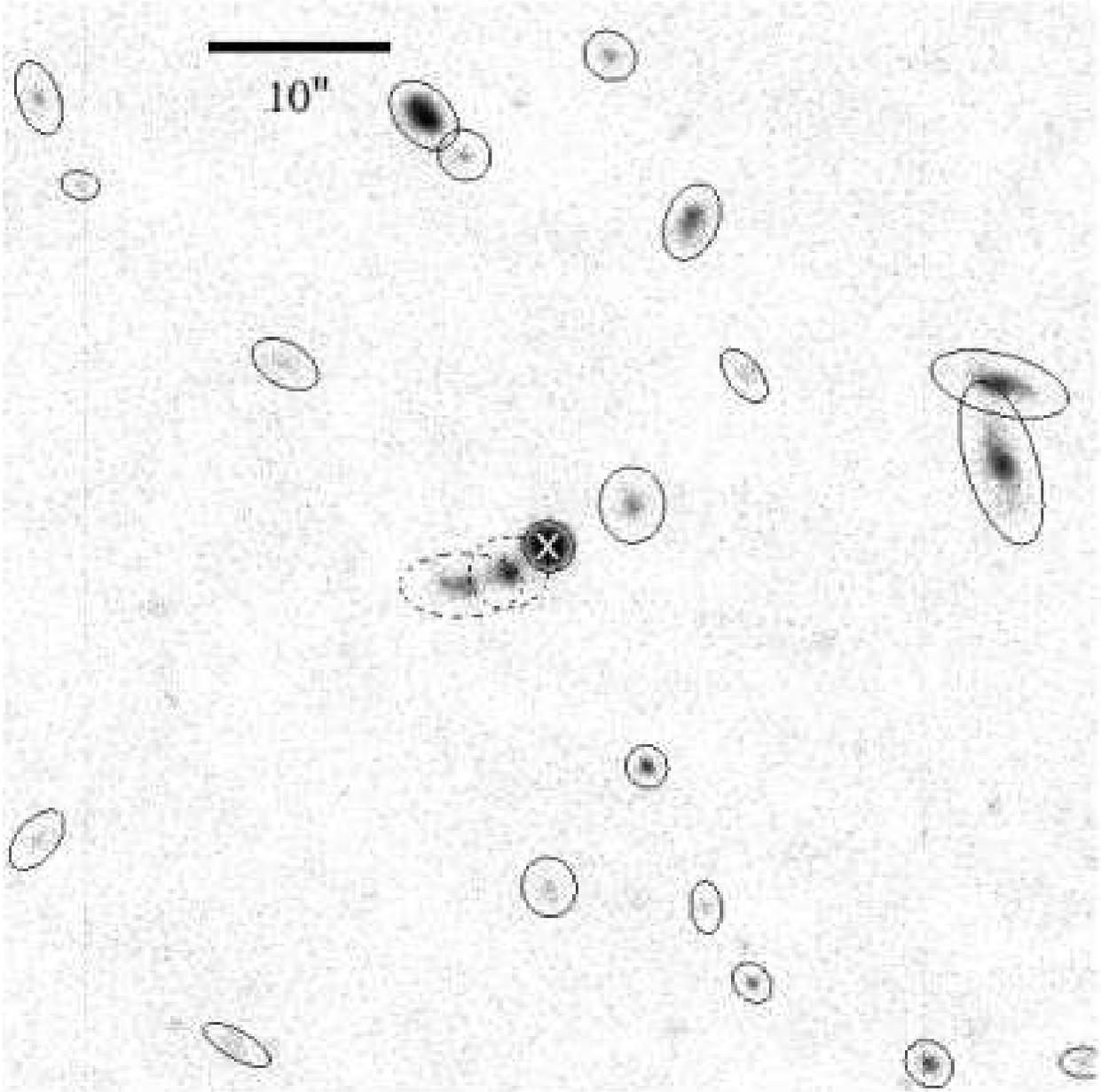}
\caption{WIYN $i^\prime$ image of the Q0902+3722 field.  The absorption
has $z_{abs}=0.6697$ and \W$=3.97$ \AA.  At the absorption
redshift, $10^{\arcsec}$ corresponds to 70.2 kpc.  The quasar PSF has
not been subtracted in this image.}
\label{image:q0902}
\end{figure}

\begin{figure}
\plotone{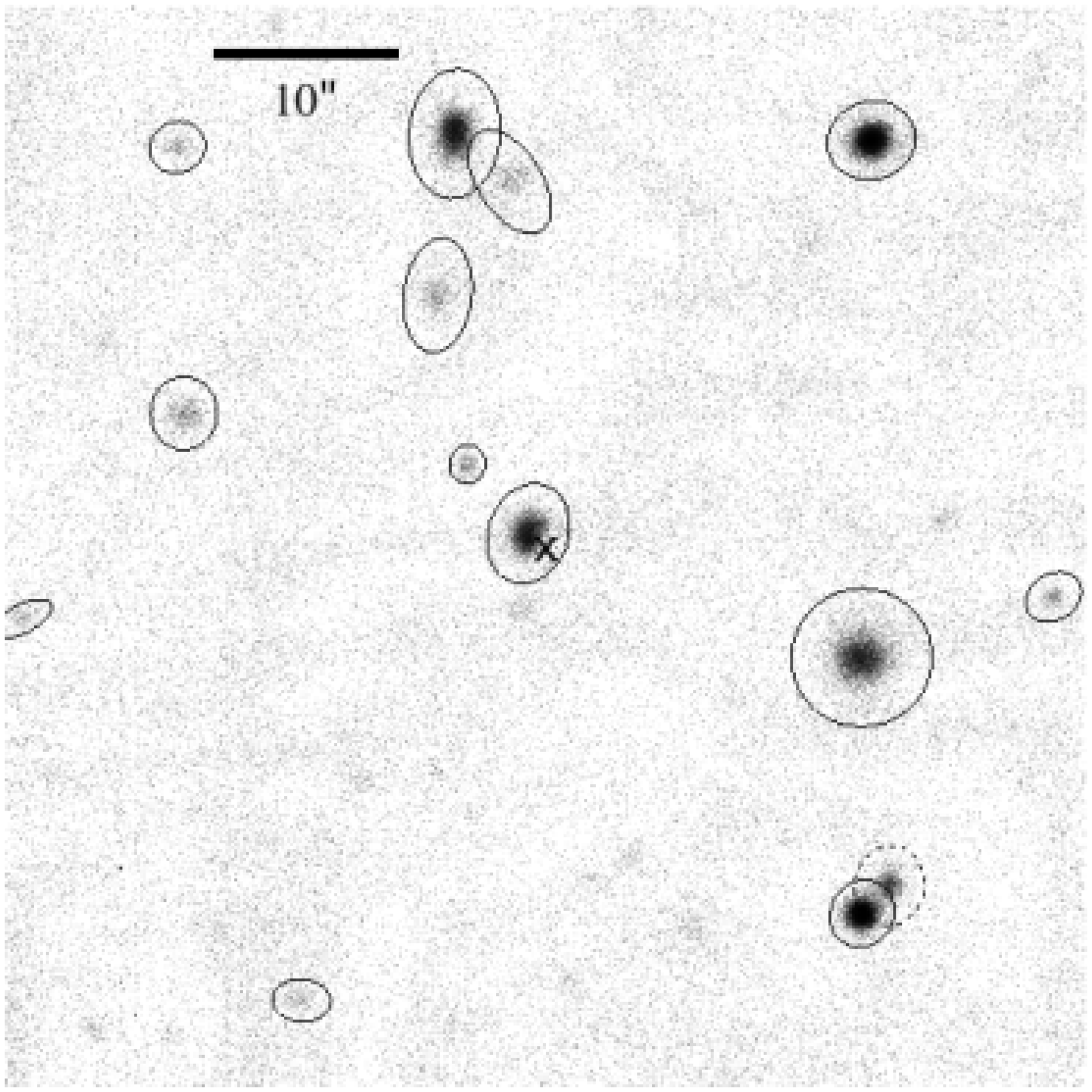}
\caption{WIYN $i^\prime$ image of the  Q1000+4438 field.  The
absorption has $z_{abs}=0.7192$ and \W$=5.33$ \AA.  At the absorption
redshift, $10^{\arcsec}$ corresponds to 72.3 kpc.}
\label{image:q1000}
\end{figure}

The spectrum of Q0747+3054 (SDSS J074707.62+305415, $z_{em}=0.974$)
contains an ultra-strong (\W$=3.63 \pm 0.06$ \AA) \ion{Mg}{2} absorber
at $z_{abs}=0.7650$.  There is also weaker \ion{Mg}{2} absorption
detected at slightly lower redshift, with $z_{abs}=0.7219$ and
\W$=0.50 \pm 0.06$, which corresponds to a velocity difference of
$\approx 7400$ km s$^{-1}$ from the ultra-strong absorber.  There were
no other low redshift \ion{Mg}{2} absorption systems detected in the
spectrum above our formal \W\ limits, though visual inspection reveals
possible \ion{Mg}{2} absorption near $\lambda \approx 4300$ \AA\ (with
$z_{abs}=0.534$ and  \W $\approx0.4$ \AA). Alternatively, it may be
due to \ion{Ca}{2} (with $z_{abs}=0.090$ and $W_0^{\lambda
3935}\approx0.5$ \AA).  Our $i^{\prime}$-band image of the field is
shown in Figure~\ref{image:q0747}.  The seeing was fairly good
(FWHM$=0.47^{\arcsec}$) and we reach a limiting surface brightness of
$i^{\prime} = 24.57$ mag arcsec$^{-2}$ to within $\approx
1.5^{\arcsec}$ of the sightline, corresponding to $\approx 11$ kpc at
$z=z_{abs}$. Subtraction of the quasar PSF reveals no evidence for
objects down to $\approx 0.8^{\arcsec}$, or $b=6$ kpc.  The two
brightest objects in the field are also detected in the SDSS images;
the galaxy $5^{\arcsec}$ (38 kpc at $z=z_{abs}$) to the northwest with
$i^{\prime} \approx 19.5$ and $g^{\prime}-i^{\prime} \approx 2.1$
(consistent with the colors of a local late-type galaxy shifted to
$z=z_{abs}$), and the galaxy $9^{\arcsec}$ (63 kpc at $z=z_{abs}$) to
the northeast with $i^{\prime} \approx 20.9$ and
$g^{\prime}-i^{\prime} \approx 3$ (consistent with the colors of a
local intermediate-type galaxy shifted to $z=z_{abs}$).  Also, a
visual inspection of Figure~\ref{image:q0747} is suggestive of one or
more galaxy groups to the north and west.

Due to the wealth of objects detected in the image and the multiple
absorbers seen in the quasar spectrum, interpretation of results for
the ultra-strong \ion{Mg}{2} absorber in this field is difficult.
Further complicating the situation is the relatively low redshift of
the quasar -- there is a possibility that some of the detected sources
are in a group that includes the quasar host at $z \approx 0.97$.
There are several reasonable explanations for the presence of the
ultra-strong absorber.  One or more of the bright galaxies could be
associated with the ultra-strong absorber.  Although galaxies this
bright are rare, so are absorbers of this strength.  Alternatively, if
the bright galaxies are in the foreground, the ultra-strong absorber
could be associated with a galaxy group consisting of several of the
other detected sources.  Finally, it is possible that one of the
fainter sources is a field galaxy at the redshift of the ultra-strong
absorber.  Spectroscopy of this field would be particularly helpful
for understanding this intriguing system.


\subsubsection{Q0747+3354}
The spectrum of Q0747+3354 (SDSS J074758.65+335432.5, $z_{em}=1.700$)
contains an ultra-strong (\W$=4.69 \pm 0.25$ \AA) \ion{Mg}{2} absorber
at $z_{abs}=0.6202$.  No other low redshift \ion{Mg}{2} absorption
system was detected.  Our $i^{\prime}$-band image is shown in
Figure~\ref{image:q0748}.  The seeing was poor (FWHM =
1.2$^{\arcsec}$), but we were able to obtain an excellent subtraction
of the quasar PSF.  We reach a limiting surface brightness of
$i^{\prime} = 24.30$ mag arcsec$^{-2}$.  The object to the west is
unresolved.  The galaxy 2$^{\arcsec}$ to the east has $i^{\prime} =
21.11$, corresponding to $b=12.8$ kpc and 1.4/1.7/2.0 $L_i^*$.  The
galaxy 5$^{\arcsec}$ to the northeast has $i^{\prime} = 23.14$, which
corresponds to $b=34$ kpc and 0.2/0.3/0.3 $L_i^*$.  Neither are
detected in the SDSS images.  It is likely that at least one of these
galaxies is associated with the absorber.


\subsubsection{Q0800+2150}
The spectrum of Q0800+2150 (SDSS J080005.3+215015.2, $z_{em}=1.790$)
contains an ultra-strong (\W$= 3.65\pm 0.16$ \AA) \ion{Mg}{2} absorber
at $z_{abs}=0.5716$.  There is also strong \ion{Mg}{2} absorption in
the spectrum at higher redshift ($z=0.994$ and $z=1.181$).  Our
$r^{\prime}$-band image obtained under fair seeing conditions (FWHM $=
0.54^{\arcsec}$) is shown in Figure~\ref{image:q0800}.  We reach a
limiting surface brightness of $r^{\prime} = 25.14$ mag arcsec$^{-2}$
to within $\approx 1.6^{\arcsec}$ of the sightline, corresponding to
$\approx 10$ kpc at $z=0.5716$.  There was not a sufficient number of
bright unsaturated stars in the field to subtract the quasar PSF.  We
detect six resolved objects within $b=100$ kpc of the sightline, none
of which were detected in the SDSS images.  The two detections with
the lowest impact parameters ($3^{\arcsec}$, or $b=19$ kpc) to the
northeast and southeast  have $r^{\prime}=22.22$ and
$r^{\prime}=24.63$, respectively, corresponding 0.7/1.0/1.2 $L_r^*$
and 0.1/0.1/0.1 $L_r^*$, respectively. It is likely that at least one
of these is associated with the ultra-strong absorber.


\subsubsection{Q0836+5132}
The spectrum of Q0836+5132 (SDSS J083618.76+513244.1, $z_{em}=1.550$)
contains an ultra-strong (\W$= 3.48\pm 0.21$ \AA) \ion{Mg}{2} absorber
at $z_{abs}=0.5666$.  No other low redshift \ion{Mg}{2} absorption
system was detected.  We obtained an $r^{\prime}$-band image under
poor seeing conditions (FWHM$=1.6^{\arcsec}$), but were able to  model
and remove the quasar PSF with only moderate residuals (see
Figure~\ref{image:q0836}.)  We reach a limiting surface brightness of
$r^{\prime} = 25.32$ mag arcsec$^{-2}$.  The PSF subtraction reveals a
galaxy 2$^{\arcsec}$ to the south of the sightline with $r^{\prime}
\approx 23.1$, which corresponds to $b=10$ kpc and 0.3/0.4/0.5
$L^*_r$.  It is not detected in the SDSS images.  The large, bright
galaxy 26$^{\arcsec}$ to the southwest has an SDSS spectroscopic
redshift of $z=0.113$.  It is likely that the faint, low impact
parameter galaxy is associated with the ultra-strong absorber.


\subsubsection{Q0902+3722}
The spectrum of Q0902+3722 (SDSS J090212.76+372208, $z_{em}=1.242$)
contains an ultra-strong (\W$= 3.97\pm 0.18$ \AA) \ion{Mg}{2} absorber
at $z_{abs}=0.6700$.  Weaker absorption is also detected at
$z_{abs}=0.932$ and $z_{abs}=1.121$ with \W$= 1.1$ \AA\ and \W$= 0.3$
\AA, respectively.  We obtained an $i^{\prime}$-band image of the
field under moderate seeing conditions (FWHM$=0.61^{\arcsec}$) that
reaches a limiting surface brightness of $i^{\prime} = 24.20$ mag
arcsec$^{-2}$ down to within $\approx$ 1.35$^{\arcsec}$ of the
sightline, corresponding to $b=9.5$ kpc (Figure~\ref{image:q0902}).
Due to strong residuals, removal of the quasar PSF did not improve the
minimum searchable impact parameter.  We detect three resolved sources
near the sightline, none of which are cataloged in the SDSS.  The
galaxies $3^{\arcsec}$ and $5^{\arcsec}$ to the southeast have
$i^{\prime} = 21.42$, and $i^{\prime} = 21.70$, respectively,
corresponding to $b=18$ kpc and 1.3/1.6/1.9 $L_i^*$ and $b=37$ kpc and
1.0/1.3/1.5 $L_i^*$, respectively.  The galaxy $5^{\arcsec}$ to the
northwest has $i^{\prime} = 22.38$ corresponding to $b=36$ kpc and
0.6/0.7/0.8 $L_i^*$.  At least one of these galaxies is likely to be
associated with the ultra-strong absorber.  The irregular appearance
of the galaxy further to the southeast suggests it is interacting with
either a satellite or the other galaxy to the southeast of the
sightline.


\subsubsection{Q1000+4438}
The spectrum of Q1000+4438 (SDSS J100015.51+443848, $z_{em}=1.862$)
contains an ultra-strong (\W$= 5.33\pm 0.19$ \AA) \ion{Mg}{2} absorber
at $z_{abs}=0.7192$.  There is also \ion{Mg}{2} absorption at
$z_{abs}=0.6964$ with \W$= 1.21\pm 0.12$ \AA, which is a velocity
difference of $\approx 4000$ km s$^{-1}$ from the ultra-strong
absorption.  We obtained an $i^{\prime}$-band image of the field with
seeing FWHM$=0.93^{\arcsec}$ that reaches a limiting surface
brightness of $i^{\prime}= 24.72$ mag arcsec$^{-2}$.  We were able to
model and remove the quasar PSF with only mild residuals, revealing a
$i^{\prime} \approx 21.5$ magnitude source $1^{\arcsec}$ from the
sightline that was blended with the light from the quasar (see
Figure~\ref{image:q1000}).  This corresponds to $b=8$ kpc and
1.5/1.9/2.2 $L^*_i$.  It is not detected in the SDSS data.  Also
detected in our image is a fainter galaxy $6^{\arcsec}$ to the
northeast with $i^{\prime}=23.94$, which corresponds to $b=44.3$ kpc
and 0.2/0.2/0.3 $L^*_i$.  It is likely that one of these galaxies is
associated with the ultra-strong absorber, while the other is
associated with the weaker, $z_{abs}=0.6964$ absorber.  Alternatively,
they may be a pair, and the galaxy associated with one of the two
absorbers is either not detected or associated with one of the larger
impact parameter galaxies in the field.

%

\subsubsection{Q1011+4451}
\begin{figure}

\plotone{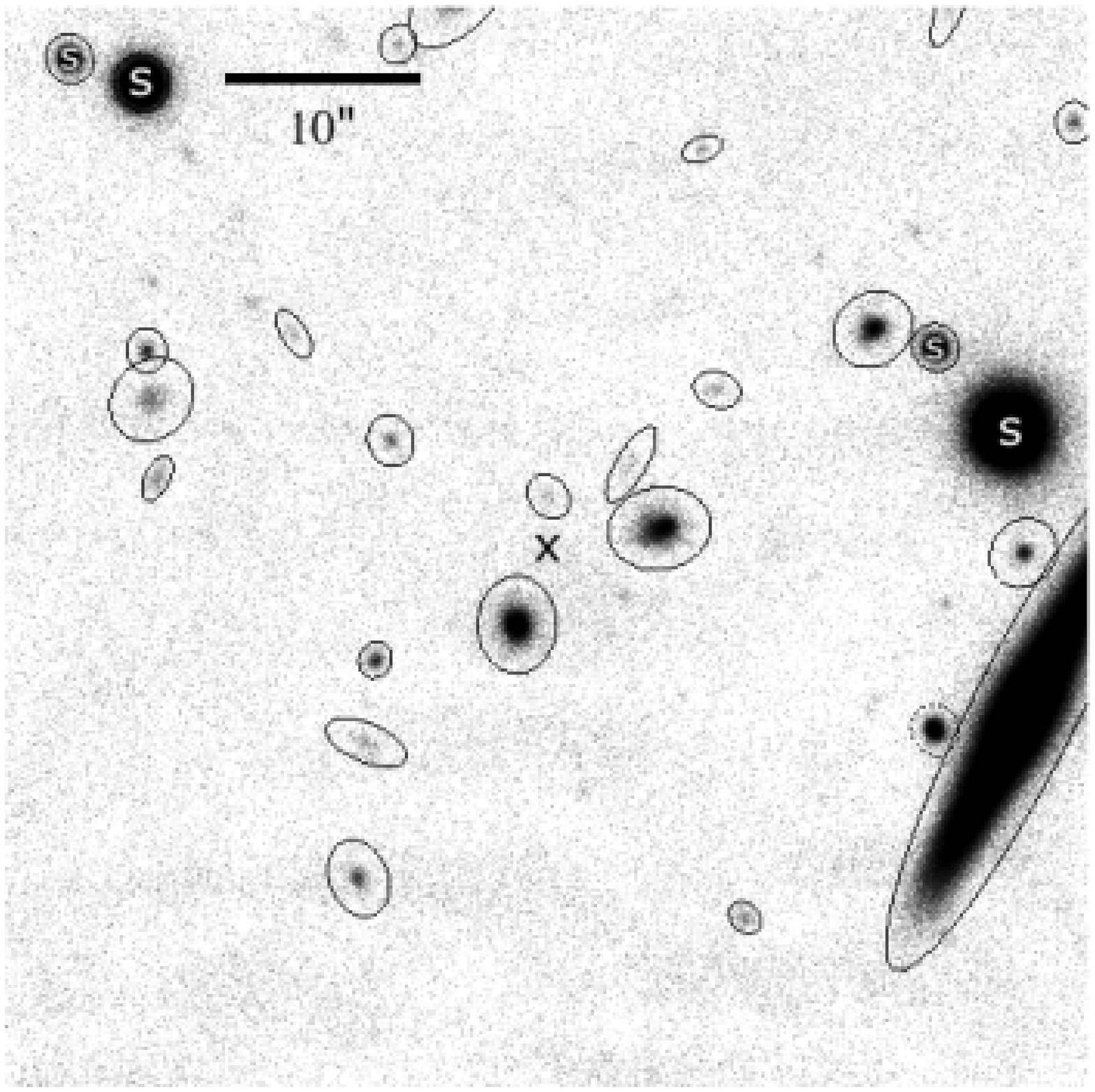}
\caption{WIYN $i^\prime$ image of the  Q1011+4451 field.  The
absorption has $z_{abs}=0.8360$ and \W$=4.94$ \AA.  At the absorption
redshift, $10^{\arcsec}$ corresponds to 76.2 kpc.}
\label{image:q1012}
\end{figure}

\begin{figure}
\plotone{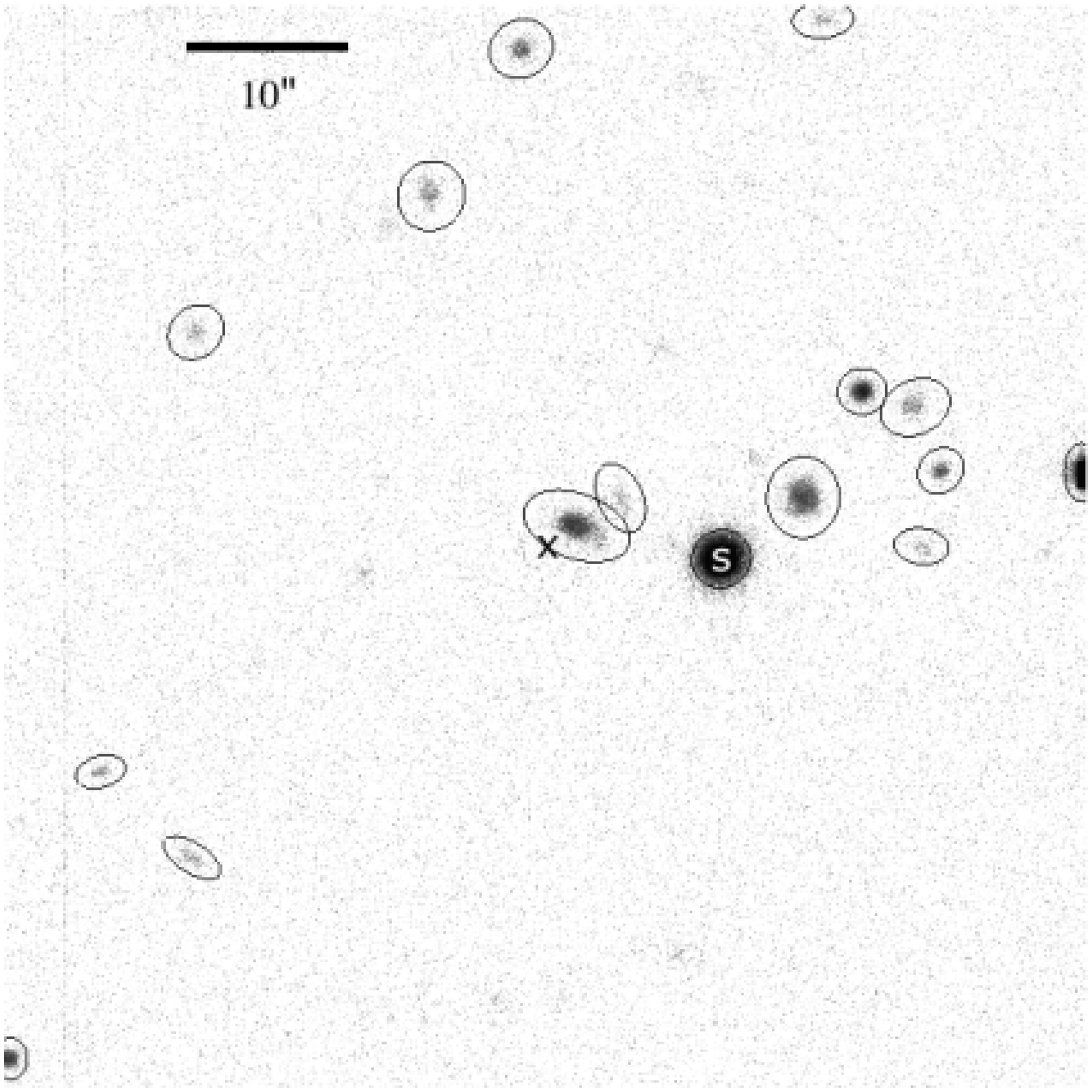}
\caption{WIYN $r^\prime$ image of the Q1038+4727 field.  The
absorption has $z_{abs}=0.5292$ and \W$=3.14$ \AA.  At the absorption
redshift, $10^{\arcsec}$ corresponds to 62.9 kpc.}
\label{image:q1038}
\end{figure}

\begin{figure}
\plotone{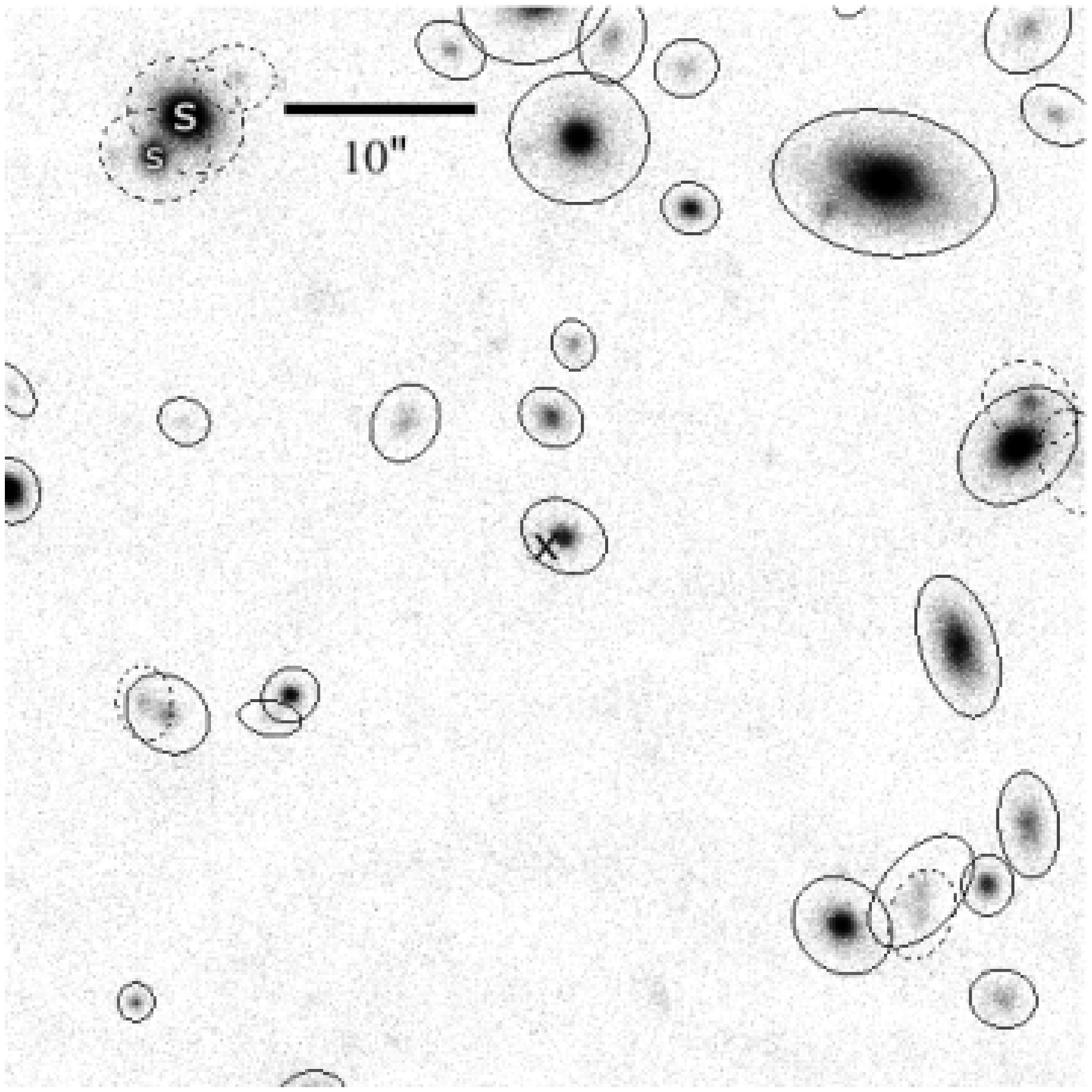}
\caption{WIYN $i^\prime$ image of the Q1356+6119 field.  The
absorption has $z_{abs}=0.7850$ and \W$=5.97$ \AA.  At the absorption
redshift, $10^{\arcsec}$ corresponds to 74.6 kpc.}
\label{image:q1356}
\end{figure}

The spectrum of Q1011+4451 (SDSS J101142.01+445155.4, $z_{em}=1.919$)
contains an ultra-strong (\W$=4.94 \pm 0.15$ \AA) \ion{Mg}{2} absorber
at $z_{abs}=0.8360$.  There is also relatively strong  \ion{Mg}{2}
absorption with \W$=1.6$ \AA\ at $z_{abs}=0.675$.  We obtained an
$i^{\prime}$-band image (Figure~\ref{image:q1012}) with a seeing of
FWHM$=0.56^{\arcsec}$ and were able to remove the quasar PSF with
moderate residuals (masked out in Figure~\ref{image:q1012}).  We reach
a limiting surface brightness of $i^{\prime}=24.60$ mag arcsec$^{-2}$.
There are two bright, low impact parameter galaxies detected in the
field.  The galaxy 4$^{\arcsec}$ to the south has $i^{\prime} =
20.63$, corresponding to $b=32$ kpc, 5.3/7.4/9.2 $L^*_i$, and a
projected optical extent of $\approx 27 \times 22$ kpc.  It is
classified as a star in the SDSS with $i^{\prime} \approx 21.1$ and
$g^{\prime}-i^{\prime} \approx 4$ (though we note that it is only
marginally detected in the SDSS data), which is consistent with the
colors of a local early-type galaxy shifted to $z=z_{abs}$.  The
galaxy 6$^{\arcsec}$ to the west has $i^{\prime} = 21.10$,
corresponding to $b=45$ kpc, 3.6/5.0/6.2 $L^*_i$, and a projected
optical extent of $\approx 30 \times 24$ kpc.  It is detected in the
SDSS images with $i^{\prime} \approx 21.4$ and $g^{\prime}-i^{\prime}
\approx 1.6$, which is consistent with the colors of a local
late-type/starburst galaxy shifted to $z=z_{abs}$.  We also detect
fainter galaxies 3$^{\arcsec}$ to the north and 6$^{\arcsec}$ to the
northwest with $i^{\prime}=24.45$ and $i^{\prime}=24.10$,
respectively, corresponding to $b=20$ kpc and 0.2/0.2/0.3 $L^*_i$ and
$b=46$ kpc and 0.2/0.3/0.4 $L^*_i$, respectively.  There are also
three other faint galaxies with $50 < b < 100$ kpc and $L < L^*_i$.
It is probable that at least one and perhaps several of the detected
galaxies are at the redshifts of the ultra-strong and strong
absorbers.  However, it would seem that no unique interpretation is
significantly more likely than the alternatives.


\subsubsection{Q1038+4727}
The spectrum of Q1038+4727 (SDSS J103808.67+472734.9, $z_{em}=1.047$)
contains an ultra-strong (\W$=3.14 \pm 0.08$ \AA) \ion{Mg}{2} absorber
at $z_{abs}=0.5292$.  No other low redshift \ion{Mg}{2} absorption
system was detected.  We obtained an $r^{\prime}$-band image
(Figure~\ref{image:q1038}) with seeing FWHM=0.82$^{\arcsec}$, reaching
a limiting surface brightness of $r^{\prime}=24.35$ mag arcsec$^{-2}$,
and were able to remove the quasar PSF with moderate residuals (masked
out in the image).  The galaxy 2$^{\arcsec}$ to the northwest has
$r^{\prime} = 21.12$, which corresponds to $b=15$ kpc and 1.5/2.1/2.4
$L_r^*$.  It is not cataloged by the SDSS.  There is also a fainter
source detected further to the northwest ($b=35$ kpc).  We consider it
likely that the low impact parameter galaxy is at the redshift of the
absorber, and perhaps part of a pair with the fainter galaxy.


\subsubsection{Q1356+6119}
The spectrum of Q1356+6119 (SDSS J135603.78+611949.6, $z_{em}=1.873$)
contains an ultra-strong (\W$=5.97 \pm 0.34$ \AA) \ion{Mg}{2} absorber
at $z_{abs}=0.7850$.  No other low redshift \ion{Mg}{2} absorber was
detected.  We obtained an $i^\prime$-band image
(Figure~\ref{image:q1356}) under moderate seeing conditions
(FWHM=0.66$^{\arcsec}$) and were able to remove the quasar PSF with
moderate residuals.  We reach a limiting surface brightness of
$i^{\prime}=24.48$ mag arcsec$^{-2}$.  The galaxy $1^{\arcsec}$ to the
northwest has $i^{\prime} \approx 21.7$, which corresponds to $b=9$
kpc and 1.7/2.2/2.7 $L_i^*$. It is likely that this galaxy is
associated with the ultra-strong absorption.  There are also three
galaxies to the north with 50 kpc $< b < 80$ kpc and $0.5L^* <
L(z=z_{abs}) < 1.5L^*$  that could potentially form a group at
$z=z_{abs}$.  None of the low impact parameter galaxies is detected in
the SDSS images.  Additionally, there may be a large group farther to
the north and west.  The presence of several bright ($i^{\prime} <
20$) galaxies suggest that they are most likely in the foreground, but
it cannot be ruled out that they are at $z \approx z_{abs}$.


\subsubsection{Q1417+0115}
\begin{figure}
\plotone{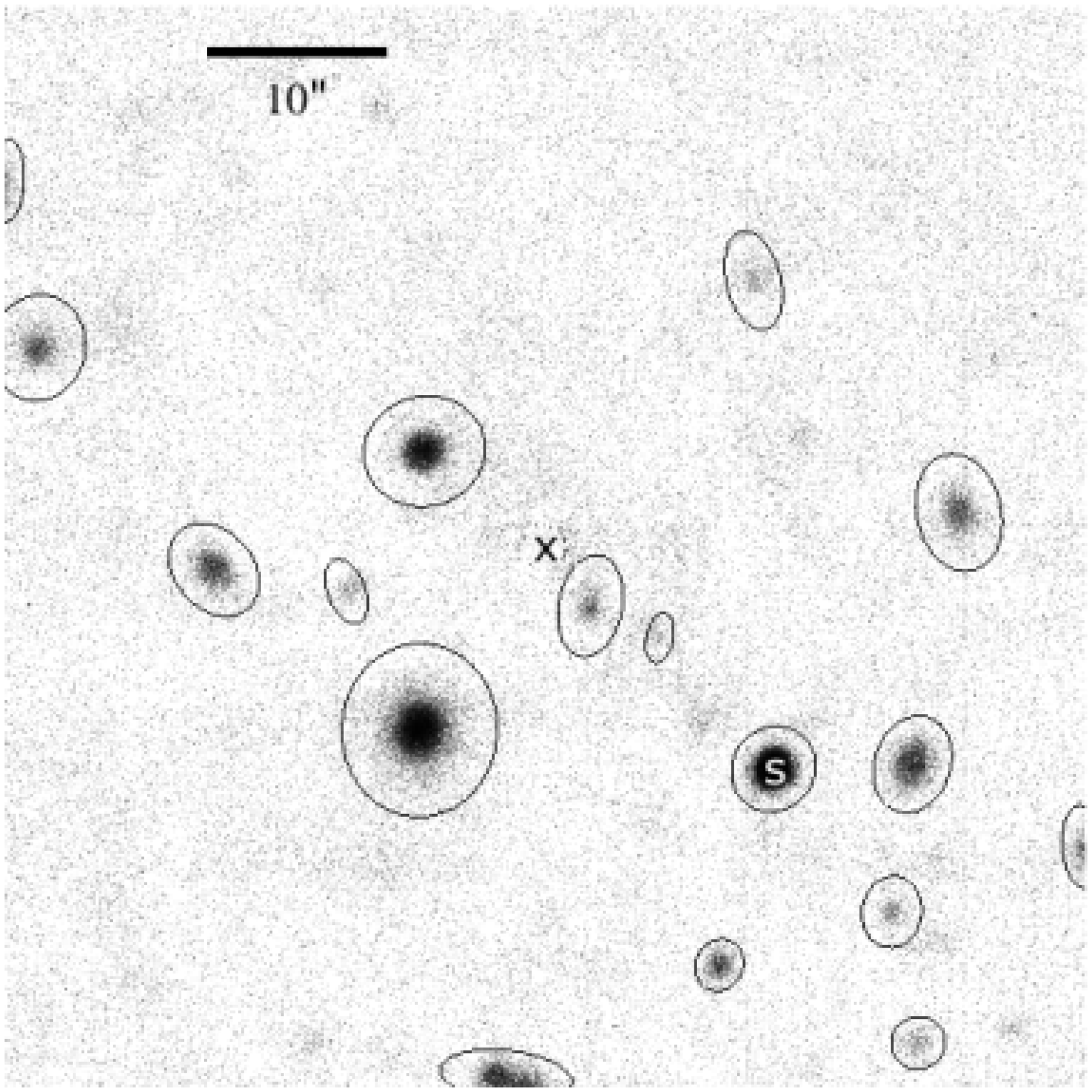}
\caption{WIYN $i^\prime$ image of the  Q1417+0115 field.  The
absorption has $z_{abs}=0.6687$ and \W$=5.6$ \AA.  At the absorption
redshift, $10^{\arcsec}$ corresponds to 70.1 kpc.}
\label{image:q1418}
\end{figure}

\begin{figure}
\plotone{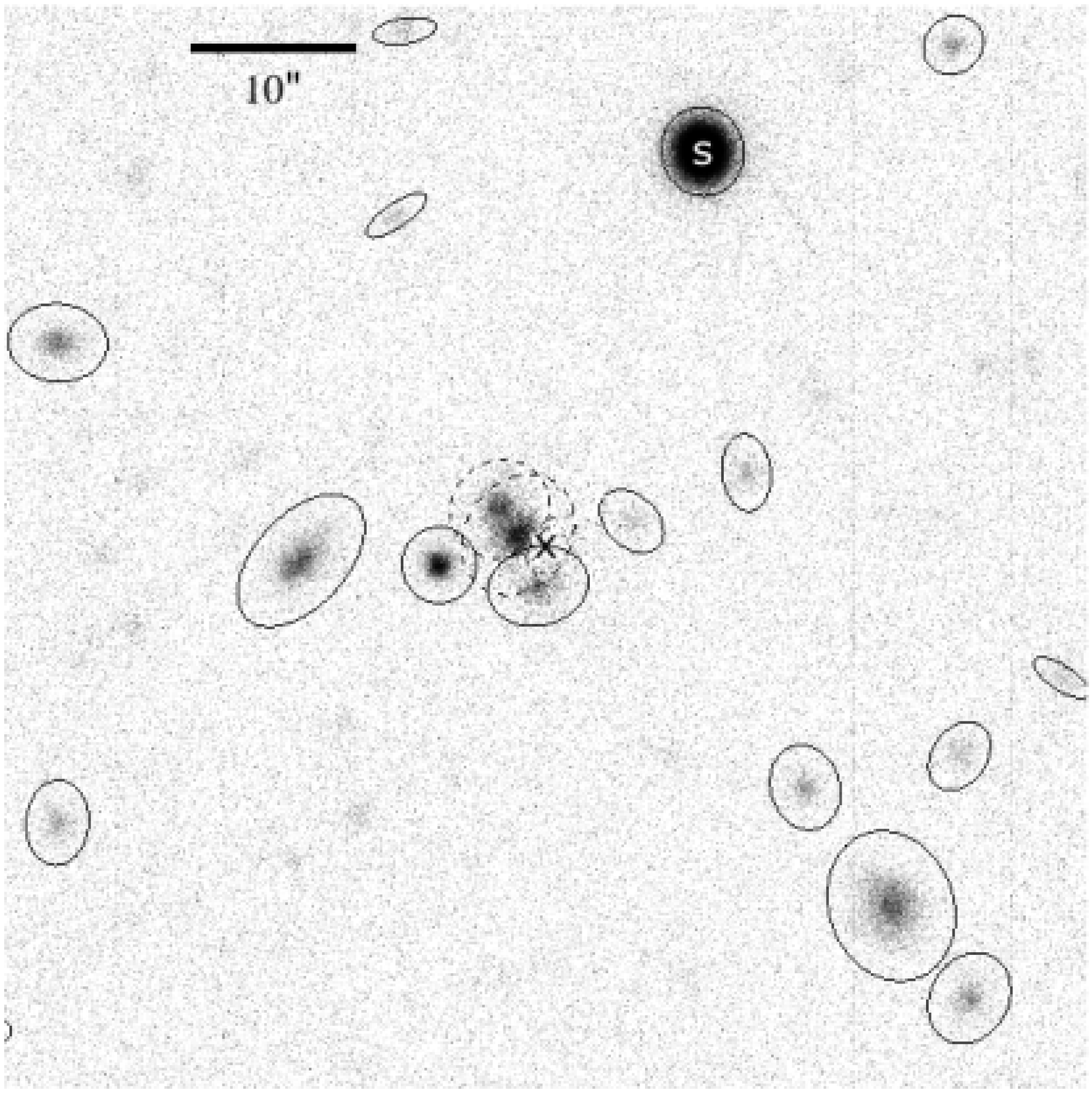}
\caption{WIYN $r^\prime$ image of the Q1427+5325 field.  The absorption
has $z_{abs}=0.5537$ and \W$=4.35$ \AA.  At the absorption redshift, 
$10^{\arcsec}$ corresponds to 64.3 kpc.}
\label{image:q1428}
\end{figure}

\begin{figure}
\plotone{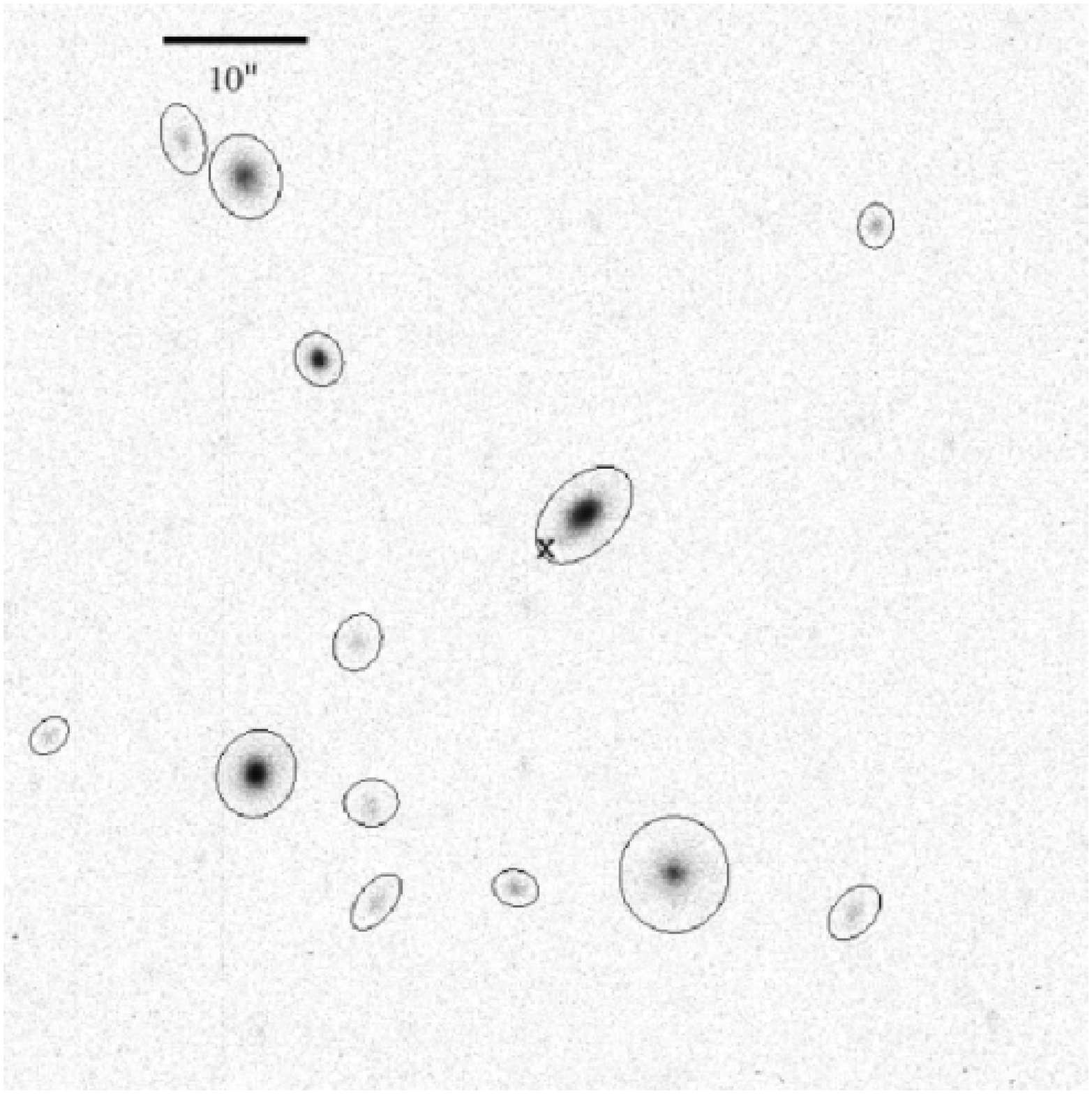}
\caption{WIYN $r^\prime$ image of the Q1520+6105 field. The absorption
has $z_{abs}=0.4235$ and \W$=4.24$ \AA.  At the absorption redshift, 
$10^{\arcsec}$ corresponds to 55.6 kpc.}
\label{image:q1520}
\end{figure}

The spectrum of Q1417+0115 (SDSS J141751.84+011556.1, $z_{em}=1.727$)
contains an ultra-strong (\W$=5.6 \pm 0.5$ \AA) \ion{Mg}{2} absorber
at $z_{abs}=0.6689$.  No other low redshift \ion{Mg}{2} absorber was
detected.  We obtained an $i^\prime$-band image
(Figure~\ref{image:q1418}) under poor seeing conditions
(FWHM=1.2$^{\arcsec}$).  The PSF in the image was asymmetric due to
wind-shake and the process of removing the quasar PSF was somewhat
uncertain, leaving strong residuals (masked out in
Figure~\ref{image:q1418}) within $r=10$ pixels ($b=7.9$ kpc).  We
reach a limiting surface brightness of $i^{\prime}=23.90$ mag
arcsec$^{-2}$.  There are two relatively bright galaxies within  100
kpc of the sightline.  The galaxy 9$^{\arcsec}$ to the northeast has
$i^{\prime}= 20.33$, which corresponds to $b=60$ kpc, 3.6/4.4/5.2
$L_i^*$, and a projected optical extent of $\approx 27 \times 25$ kpc.
It is classified as a star by the SDSS with $i^{\prime} \approx 20.6$
and $g^{\prime}-i^{\prime} \approx 1.5$; however, the colors are
consistent with those of a local late-type/starburst galaxy shifted to
$z=z_{abs}$.  The galaxy 12$^{\arcsec}$ to the southeast has
$i^{\prime}= 19.32$, which corresponds to $b=86$ kpc, 9.1/11.0/13.2
$L_i^*$, and a projected optical extent of $\approx 42 \times 38$ kpc.
It is detected in the SDSS images with $i^{\prime} \approx 18.9$ and
$g^{\prime}-i^{\prime} \approx 2.4$, which is consistent with the
colors of a local intermediate/late-type galaxy shifted to
$z=z_{abs}$.  There are also fainter resolved sources detected in our
image including a low impact parameter galaxy 4$^{\arcsec}$ ($b=29$
kpc) to the southwest with $i^{\prime}= 21.86$ (0.8/1.0/1.2 $L_i^*$).
This low impact parameter galaxy may be associated with the
ultra-strong absorption, while the bright galaxies are in the
foreground.  However, it is possible that the bright galaxies are
indeed (also) at $z=z_{abs}$, and are part of a group or exhibit
superwinds, accounting for the ultra-strong \W\ of this absorber.


\subsubsection{Q1427+5325}
The spectrum of Q1427+5325 (SDSS J142749.37+532508.9, $z_{em}=0.909$)
contains an ultra-strong (\W\ $=4.35 \pm 0.17$ \AA) \ion{Mg}{2}
absorber at $z_{abs}=0.5537$.  No other low redshift \ion{Mg}{2}
absorber was detected.  We obtained an $r^\prime$-band image
(Figure~\ref{image:q1428}) under poor seeing conditions
(FWHM=1.1$^{\arcsec}$).  We were able to remove the quasar PSF, but
with moderately strong residuals.  We reach a limiting surface
brightness of $r^{\prime} = 24.81$ mag arcsec$^{-2}$.  There appear to
be three separate sources blended with the quasar PSF, although the
poor seeing and residuals make it difficult to determine if they are
in fact separate sources or if some of them are the same object.  They
are not cataloged in the SDSS.  If at $z=z_{abs}$ they would total $L
\approx$ 3-4$L^*$ within $b \le 35$ kpc.  There are also several
additional galaxies with impact parameters and luminosities in the
range 35 kpc $<b<$ 95 kpc and  $0.1L^*_r \lesssim L(z=z_{abs})
\lesssim 1.5L^*$.  We consider it very likely that the galaxy (or
galaxies) surrounding the sightline are associated with the
ultra-strong absorption, possibly  forming part of a larger group with
the other nearby galaxies.  We also note, however, that due to the
relatively low redshift of the quasar it is possible that we are
detecting some light from the quasar host as well.


\subsubsection{Q1520+6105}
The spectrum of Q1520+6105 (SDSS J152046.36+610511.3, $z_{em}=2.182$)
contains an ultra-strong (\W\ $=4.24 \pm 0.14$ \AA) \ion{Mg}{2}
absorber at $z_{abs}=0.4235$.  No other low redshift \ion{Mg}{2}
absorber was detected.  We obtained an $r^\prime$-band image
(Figure~\ref{image:q1520}) in fair seeing conditions
(FWHM=0.76$^{\arcsec}$) and were able to remove the quasar PSF.  We
reach a limiting surface brightness of $r^{\prime} = 24.03$ mag
arcsec$^{-2}$.  The galaxy 4$^{\arcsec}$ northwest of the sightline
has $r^{\prime} = 19.72$, which corresponds to $b=20$ kpc and
2.9/3.5/3.9 $L_r^*$. It is the only source that we detect within $b
\le 80$ kpc of the sightline and consider it likely that it is
physically associated with the ultra-strong absorber.  It is detected
in the SDSS images with $r^{\prime} \approx  19.6$ and
$g^{\prime}-r^{\prime} \approx 1.6$, which is consistent with the
colors of a local intermediate/late-type galaxy shifted to $z=z_{abs}$.


\subsection{Statistical Properties of the Sample} 
\begin{figure}
\plotone{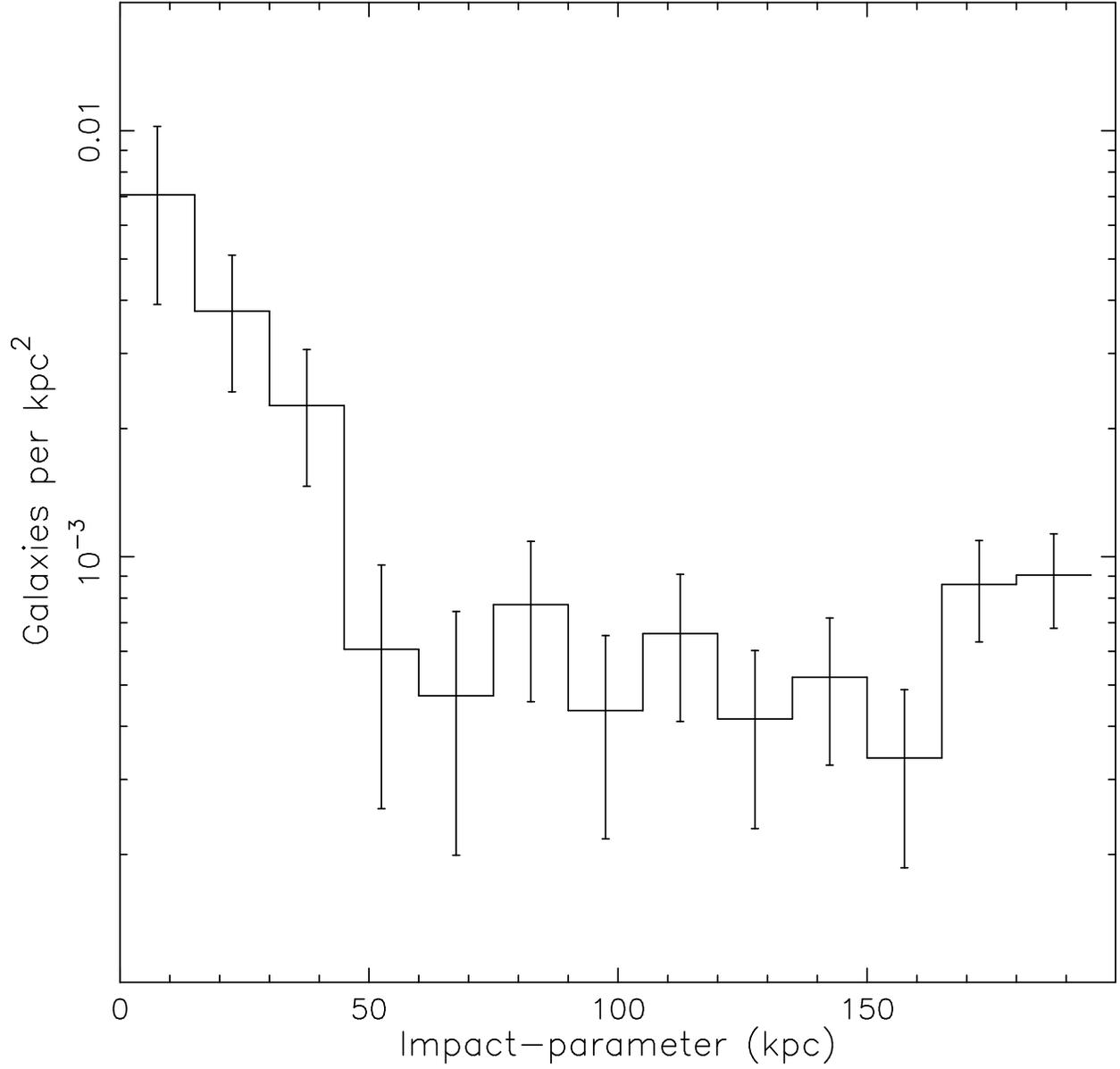}
\caption{The surface-density of galaxies with $L(z=z_{abs}) \ge
0.5L^*$ in our fields as a function of impact parameter.  There is a
clear overdensity of galaxies with $b \lesssim 45$ kpc.}
\label{gal_dens}
\end{figure}

Each of the sightlines imaged in this work contains a very strong
\ion{Mg}{2} absorption system in the redshift range $0.42 < z_{abs} <
0.84$.  Thus, it is expected that the fields surrounding each
sightline will, on average, exhibit more sources at $z \approx
z_{abs}$, in comparison to a randomly selected field.
Figure~\ref{gal_dens} demonstrates this excess along our sightlines
for the detected galaxies which would have $L(z=z_{abs}) \ge 0.5 L^*$.
There is a clear overdensity of galaxies within $b \approx 45$ kpc in
comparison to larger impact parameters by a factor of $\approx$ 4 for
30 kpc $< b < 45$ kpc and rising to a factor of $\approx$ 10 for $b <
15$ kpc.  Of course, each field is contaminated by both foreground and
background sources that are physically unrelated to the absorption
environment.  In particular, many of the apparently bright sources may
be foreground galaxies.

Our 15 fields contain a total of 13 sources that, if at $z=z_{abs}$,
would have $4 L^* \lesssim L \lesssim 13 L^*$ and $b<200$ kpc.  More
than half of these would have $b<90$ kpc, even though this covers only
$\approx20$\% of the area.  Locally, galaxies as bright as $4 L^*$ are
less numerous than $L^*$ galaxies by a factor of $\approx 25$ (Blanton
et al.\ 2003), and are almost exclusively early-type (Nakamura et. al,
2003), and/or cD cluster galaxies (e.g., Laine et al.\ 2003), while
galaxies with $L \gtrsim 8 L^*$ are extremely rare, having number
densities $\lesssim 10^{-3}$ times that for $L^*$
galaxies.\footnote{We note, however, that bright galaxies were more
common in the past.  The results of Gabasch et al., 2006 suggest that
$4 L^*$ galaxies were more common at $0.45 < z < 0.85$ by a factor of
$\approx 2.5$ and $8 L^*$ galaxies by a factor of $\approx 20$,
compared to the Blanton et al.\ 2003 results.}  Thus, it is possible
that some of them are in the foreground of the absorber, and
consequently less luminous than the $z_{gal}=z_{abs}$ assumption
implies.  In order to explore whether all of the bright galaxies in
our sample can be accounted for by foreground objects, we created
simulated galaxy catalogs for the fields under study using the galaxy
number counts in the $r^{\prime}$ and $i^{\prime}$ bands from the SDSS
commissioning data (Yasuda et al.\ 2001).  The counts in that study
reliably reach at least $r^{\prime} \approx i^{\prime} \approx 20$.
Beyond this value, we extrapolate using the faint-end of the
number-magnitude curve. Subaru Deep Field data (Nagashima et al.\
2002) show that the counts-magnitude relation turns over rather slowly
at fainter visual magnitudes.  By extrapolating from the Yasuda et
al.\ results, we at worst slightly overpredict the expected number of
galaxies at the faint end of our modeled sample, but this has no
effect on  the predicted number of bright galaxies.

\begin{figure}
\plotone{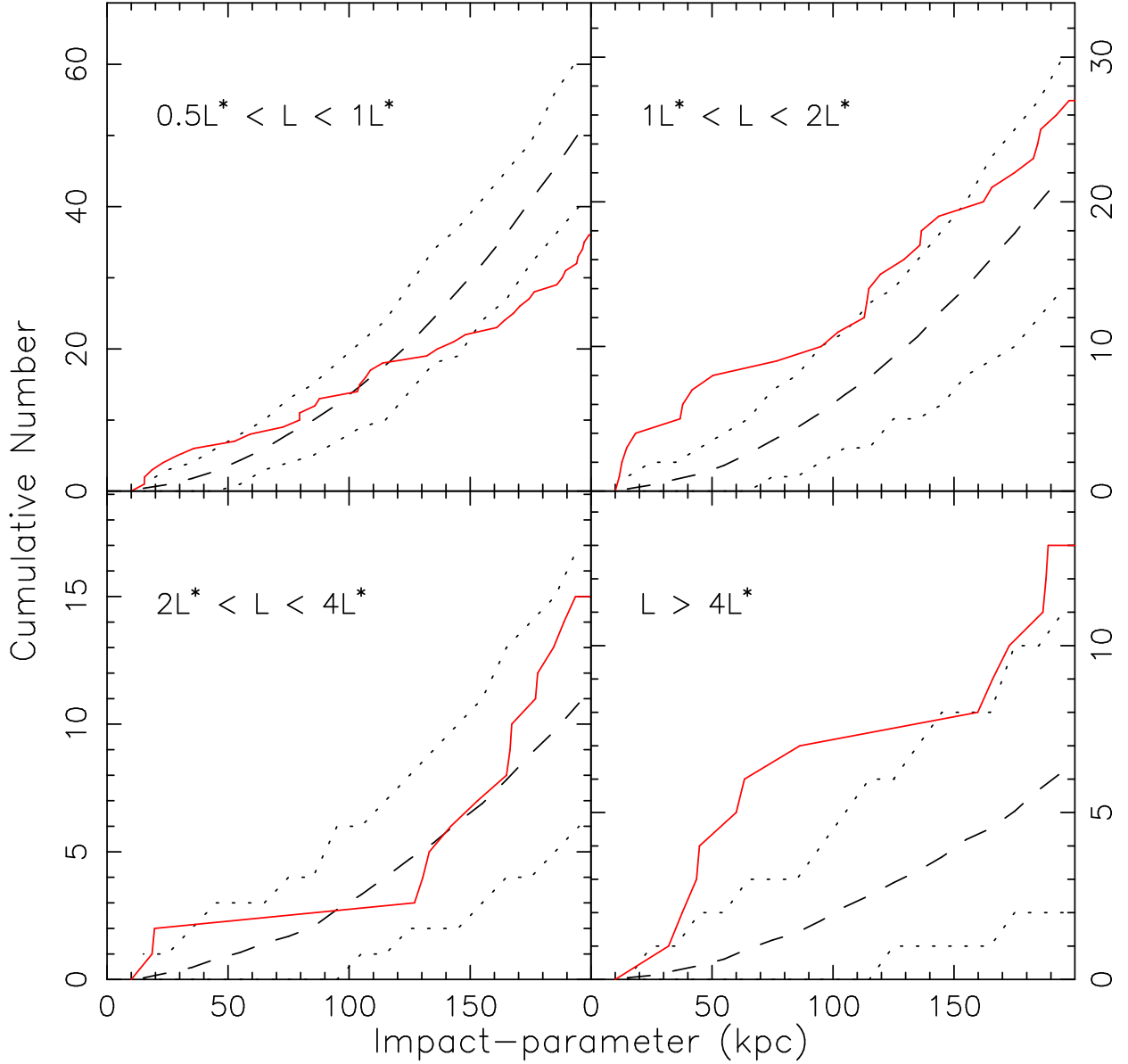}
\caption{Cumulative distribution of galaxies as a function of impact
parameter.  The solid lines indicate the data.  The dashed lines are
the averages of 100 Monte Carlo simulations, and the dotted lines
indicates the ranges containing 95\% of the simulations.  The
excess/dearth of galaxies at any impact-parameter relative to the
simulation can be seen by comparing the slopes of the curves.  There
is an excess of galaxies at low impact parameter for sources with 
$L(z=z_{abs}) \ge 0.5 L^*$.  The excess is particularly significant
for $L(z=z_{abs}) \ge 4 L^*$. }
\label{gal_numbers}
\end{figure}

We randomly populated each field with galaxies according to the
number-magnitude relation for 100 Monte Carlo simulations.  For
various luminosity ranges (assuming all have $z_{gal}=z_{abs}$ and
using the average of Sa- and Sc-type $k$-corrections) we then computed
the cumulative number of galaxies within a given impact parameter for
the sum of all sightlines, and we compared these results to our data.
The results are shown in Figure~\ref{gal_numbers}.  The upper-left
panel shows the cumulative distribution for galaxies with $0.5 L^* \le
L(z=z_{abs}) < 1 L^*$.  We find fewer galaxies within $b=200$ kpc than
expected from our simulation at $> 95$\%  confidence.  While cosmic
variance, which is not considered in our simulation, would not affect
the overall average, it would expand the range of the confidence
intervals, and thus this dearth could be due to cosmic variance.  More
likely, however, it is due to an overestimation of the number of
apparently-faint galaxies in our simulation due to the inexact
extrapolation of the faint end of the number-magnitude relation.
Notably, there is a significant {\it excess} of galaxies within $b
\approx 60$ kpc.  The low impact parameter galaxy excess is seen more
distinctly in the upper-right panel which shows the distribution for
galaxies with $1 L^* \le L(z=z_{abs}) < 2 L^*$.  Surprisingly,
although the excess is less significant for galaxies with $2 L^* \le
L(z=z_{abs}) < 4 L^*$, it is {\it very} significant for galaxies with
$L(z=z_{abs}) \ge 4 L^*$.

\begin{figure}
\plotone{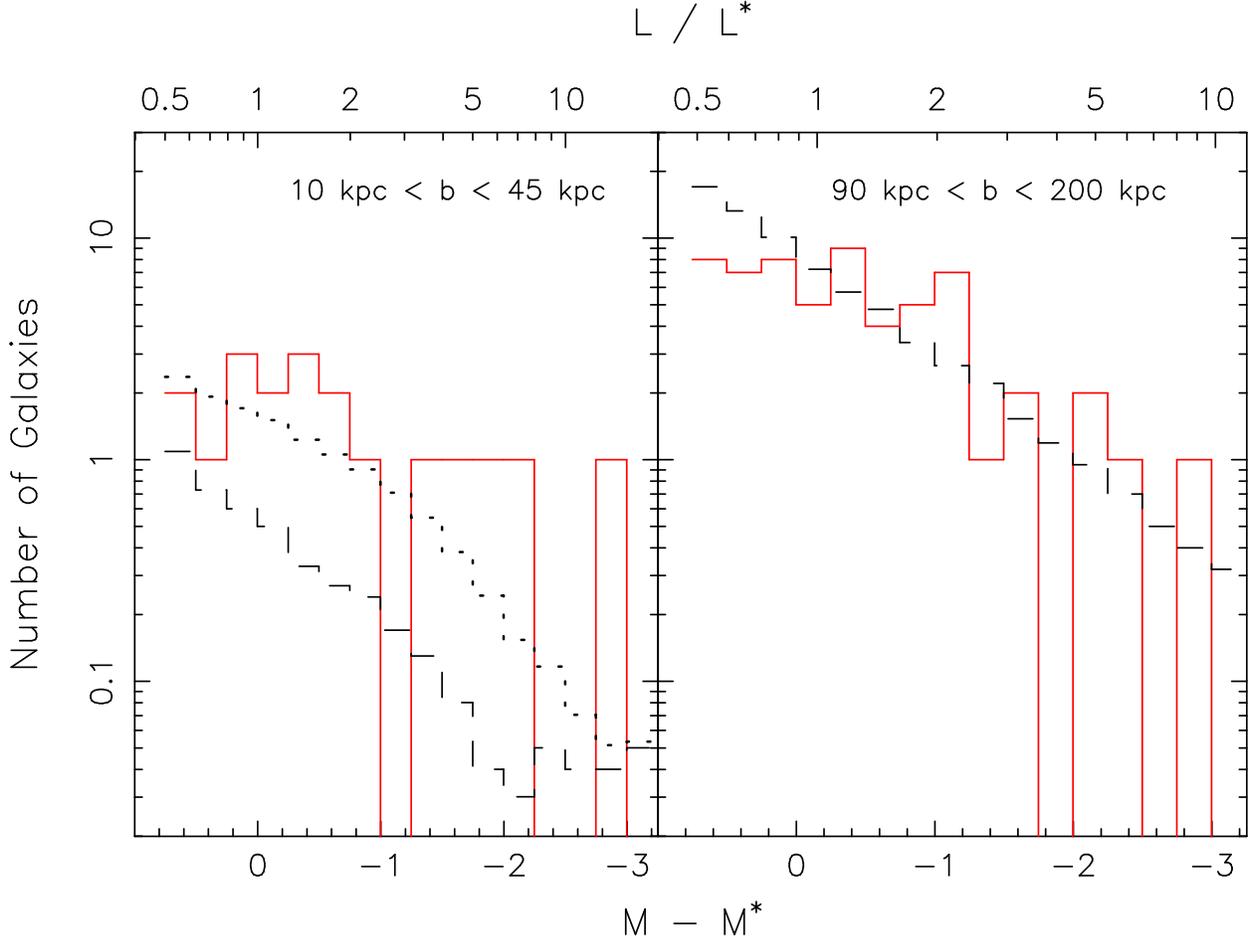}
\caption{Distribution of galaxies having $L(z=z_{abs}) \ge 0.5L^*$ as
a function of magnitude.  The solid lines indicate the data.  The
dashed lines are the averages of 100 Monte Carlo simulations.  Left:
galaxies with impact parameters 10 kpc $< b <$ 45 kpc.  There is a
clear excess of galaxies compared to the simulations.  The dotted line
is the sum of the simulation result and the renormalized
$r^{\prime}$-band LF at $0.45 < z \le 0.85$ from Gabasch et
al.\ (2006).  Right: galaxies with impact parameters  90 kpc $< b <$
200 kpc.  There is good agreement between the observed galaxy
number-magnitude relation and the simulations, indicating that any
correlation between absorbers and galaxies with $b > 90$ kpc is weaker
than we can detect with our current sample.}
\label{gal_LF}
\end{figure}

In Figure~\ref{gal_LF}, we show the number of galaxies with $L \ge 0.5
L^*$ as function of magnitude for two ranges of impact parameter, 10
kpc $< b < $ 45 kpc and 90 kpc $< b < $ 200 kpc, as well as the
predictions from the Monte Carlo simulations.  The excess at low-$b$
can be seen to hold over a large range of luminosity.  For $b > 90$
kpc, there is good agreement between the measured galaxy
number-luminosity relation and the simulations.  Thus, the larger-$b$
galaxies are consistent with being uncorrelated with the presence of
the ultra-strong \ion{Mg}{2} absorption at the level detectable by our
small sample.  The excess at low-$b$ does not necessarily represent
the LF of ultra-strong \ion{Mg}{2} absorber {\it galaxies}, but rather
the LF of all galaxies in ultra-strong \ion{Mg}{2} absorber {\it
environments}.  However, there are 19 detected sources in this
magnitude and impact-parameter range, compared to an average of 4.4 in
our simulations.  Although this does not include sources fainter than
$0.5L^*$ or within $b \le 10$ kpc (which total an additional 14
sources), the net average number of galaxies over the expected
foreground plus background is approximately one per sightline.

To compare this excess to the LF of field galaxies at similar
redshift, we also show in Figure~\ref{gal_LF} the results of the
simulation plus the $r^{\prime}$-band LF at $0.45 < z \le 0.85$ from
Gabasch et al.\ (2006) renormalized (by a factor of $\simeq 40$) such
that the sum matches the data in the $0.5 L^* < L < 1.0 L^*$ range.
The shape of the LF for  ultra-strong \ion{Mg}{2} absorber
environments at $0.4 \lesssim z \lesssim 0.8$ appears to be roughly
consistent with that of random environments at similar redshift, with
perhaps an excess of galaxies with $1 L^* \le L < 2 L^*$ relative to
slightly fainter and slightly brighter galaxies.  Larger samples are
necessary to confirm/quantify this.  Additionally, even though bright
galaxies were more common in this redshift range than at $z=0$, the
expected number of $L > 4 L^*$ galaxies with 10 kpc $< b < 45$ kpc
from the sum of the Monte Carlo simulations plus the $0.45 < z \le
0.85$ LF is only $\left<n_{gal}\right> \simeq 1$, whereas we detected
four.

The excess of galaxies with $4 L^* \le L(z=z_{abs}) < 13 L^*$ seen in
Figures~\ref{gal_numbers} and~\ref{gal_LF} is unlikely to be due to
cosmic variance.  While a chance low redshift galaxy group along one
of our sightlines could produce an apparent excess, we note that the
putative $4 L^* \le L < 13 L^*$ galaxies are distributed among several
fields, making the idea of appealing to some kind of foreground effect
less likely.  In our data, the seven galaxies with $4 L^* \le
L(z=z_{abs}) < 13 L^*$ and $b<90$ kpc are distributed over four
fields: Q0013+1414, with $L(z=z_{abs}) \approx 5 L^*, b=44$ kpc;
Q0747+3054 with $L(z=z_{abs}) \approx 12 L^*, b=38$ kpc and
$L(z=z_{abs}) \approx 4.5 L^*, b=63$ kpc; Q1011+4451 with
$L(z=z_{abs}) \approx  6.4 L^*, b=32$ kpc and $L(z=z_{abs}) \approx 4
L^*, b=45$ kpc; and Q1417+0115 with $L(z=z_{abs}) \approx 4 L^*, b=60$
kpc and $L(z=z_{abs}) \approx 10 L^*, b=86$ kpc.  The six $4 \le
L(z=z_{abs}) < 13 L^*$ galaxies with $b>90$ kpc are all in other
fields, with one each towards Q0800+2150 and Q0902+3722 and four
towards Q1356+6119. We also note that all of these larger impact
parameter bright galaxies have $b \ge 160$ kpc, i.e., there is a gap
from $b=86$ kpc to $b=160$ kpc where we detect no galaxies brighter
than $L(z=z_{abs}) \approx 4 L^*$.  Thus, while cosmic variance cannot
be ruled out as the explanation for some of the bright, low impact
parameter galaxies in our data, it seems likely that the bulk are at
the absorption redshift and in some way related to the very large
velocity spreads of the \ion{Mg}{2} absorption along the quasar
sightlines.

\section{Discussion}
\label{section:discussion}
Each of our fields contains at least one detected galaxy with $L
\gtrsim 0.3 L^*$ and $b \lesssim 40$ kpc, and many have detected
galaxies with optical extents blended with the quasar PSF.  In studies
of galaxies associated with quasar absorption lines, it is customary
to simply state luminosities and impact parameters.  However, when the
impact parameter is small, the {\it size} of the galaxy is also
relevant.  More specifically, it is of interest to know impact
parameters scaled by some measure of the luminous radii of the
galaxies in the field.  Therefore, we define the radius of the optical
extent of the galaxy along the direction towards the quasar,
$R_{gal-quasar}$, and consider the ratio of the impact parameter $b$
to this quantity, $\varepsilon = b/R_{gal-quasar}$.  Though the
measured values of $R_{gal-quasar}$ will be sensitive to the galaxy's
luminosity and redshift, as well as the seeing quality and depth of
our photometry in each field, $\varepsilon$ is nonetheless suggestive
of a more physical, if approximate, indication of the region of a
galaxy that is being sampled by the sightline.  Using this
measurement, we find that five of our fields (Q0232-0811, Q0747+3354,
Q1000+4438, Q1356+6119, and Q1427+5325) have a galaxy overlapping the
sightline (i.e., $\varepsilon < 1$).  Two additional fields
(Q0836+5132 and Q1520+6105) have detected galaxies with $\varepsilon
\lesssim 1.3$.  In seven of the remaining fields the closest detected
galaxies have $1.5 < \varepsilon < 2.5$ and one field (Q1417+0115) has
$\varepsilon \approx 4$.  Many of the low impact parameter galaxies
detected in our sample appear to have a fainter neighbor.  We quantify
this in a similar fashion as above.  If a galaxy with $b<50$ kpc has a
neighbor within $\varepsilon_{gal-gal} < 2$ we consider this evidence
for a ``pair''.  Five of our 15 sightlines meet this criterion, and
another three have $2 < \varepsilon_{gal-gal} < 3$, where we list the
evidence for a pair as ``close.''  Using this information, together
with the images themselves, we can tentatively categorize each field
as ``normal'' (i.e., similar to sightlines of less-strong \ion{Mg}{2}
absorbers, with a $0.3L^* \lesssim L \lesssim 2L^*$ galaxy at $b
\lesssim 50$ kpc), ``overlap'' (i.e., having sightlines which
intersect the optically-luminous extents of the galaxies),
``interacting'' (i.e., exhibiting evidence for distorted morphology or
interacting galaxies), and/or ``bright'' (i.e., having multiple $L
\gtrsim 4 L^*$ galaxies with $b \le 90$ kpc in the field.)  We
summarize these findings in Table~\ref{table:environment}.  We also
note that visual inspection of the images suggests that a cluster or a
large group may be present in several of the fields, particularly
Q0747+3054, Q0800+2150, and Q1356+6119.

Our four weakest systems, which have 2.7 \AA\ $\lesssim$\W$\lesssim$
3.5 \AA, are all categorized as ``normal''.  The seven systems with
3.6 \AA\ $\lesssim$\W$\lesssim$ 4.7 \AA\ are represented by all of our
categories.  Among the four strongest absorbers, which have 5 \AA\
$\lesssim$\W$\lesssim$ 6 \AA, two are categorized as ``bright'' and
two as ``overlap''.  However, we emphasize that the number of
sightlines we have studied is small, and that apparent trends with \W\
may not be supported by larger samples.

Surprisingly, almost half of the images in our sample appear similar
to fields selected by less-strong \ion{Mg}{2} absorbers, i.e., in the
``normal'' category.  Thus, we have little additional evidence to
diagnose the cause of the large kinematics seen in absorption for
those systems.  It is therefore possible that some ultra-strong
\ion{Mg}{2} systems simply represent the tail of the distribution of
galaxy kinematics and chance alignment of the sightline with many
low-ionization absorbing clouds in the galaxy halo. This
interpretation is consistent with the findings  of Rao et al.\ (2006)
where they find that, while the fraction of \ion{Mg}{2} absorbers that
are DLAs increases with increasing \W, not all ultra-strong
\ion{Mg}{2} systems are DLAs.  In this picture, an increasing number
of kinematically-distinct ``clouds'' intercepted by the sightline
increases both the observed \W\ and the likelihood that at least one
of the clouds exhibits a \ion{H}{1} column density above the DLA
threshold.

Five to seven of our images reveal the presence of a galaxy
overlapping (or nearly so) the sightline.  This raises the question of
whether many ultra-strong systems might simply select sightlines
through the inner regions (as opposed to the extended-halos) of normal
galaxies.  However, Galactic \ion{Mg}{2} absorption seen against
quasars has only a median $W_0^{\lambda 2796} = 1.17 \pm 0.33$ \AA\
(Savage et al.\ 2000) which corresponds to $\Delta V_{REW} = 125 \pm
35$ km s$^{-1}$.  These systems, of course, sample only roughly half
of a complete sightline through the Galactic disk.  Nonetheless, the
absorption kinematics tend to be at most only slightly asymmetric
about the Local Standard of Rest.   To account for the asymmetry, we
use the minimum/maximum velocity limits of each absorber given by
Savage et al.\ to estimate that a full line of sight through the disk
would only increase the observed absorption strengths by $\approx$
10\% to 50\%, and in no case more than $\approx$70\%.  Thus, it is not
expected that the Milky Way would produce an ultra-strong \ion{Mg}{2}
absorber, even with a sightline passing through the entire disk at the
solar locus.  If there is a significant intermediate redshift
population with smaller velocity spreads (i.e., smaller
$W_0^{\lambda2796}$ values) but having similar overlapping
galaxy--sightline pairs, the discrepancy would simply be due to our
observing only ultra-strong systems.  The absence of this population,
however, would mean that our sample is fairly representative and would
imply that galaxies containing significant amounts of neutral/low-ion
gas at $0.4 \lesssim z \lesssim 0.8$ were in general kinematically
more complex than the Milky Way.  This is consistent with the findings
of NTR05 whereby the total proper cross section for absorption by
ultra-strong \ion{Mg}{2} absorbers is decreasing with decreasing
redshift, especially for $z \lesssim 1$, indicating a larger incidence
of ultra-strong systems at intermediate redshift compared to the
present epoch.  We also note that  this difference is consistent with
the decreasing global star formation rate from $z \approx 1$, which
may be relevant if the kinematics are driven by star formation.

Using HST ACS images, Kacprzak, Churchill, \& Steidel (2005) have
argued that, for systems with 0.03 \AA\ $\lesssim$ \W\ $\lesssim 1.2$
\AA, the absorption-line kinematics are correlated with galaxy
asymmetry divided by impact parameter.  Of course, our \W\ values are
significantly larger and we cannot produce their asymmetry
measurements for our sample with ground-based data; however, it is
interesting that at least a third of our sample shows evidence for a
low impact parameter pair.  Although some of these putative pairs are
likely foreground or background coincidences, the presence of an
interacting pair could provide the kinematic complexity necessary for
the presence of an ultra-strong \ion{Mg}{2} absorber.

In addition, the evidence for $4 L^* \lesssim L \lesssim 13 L^*$
galaxies with $b \le 90$ kpc has interesting consequences.  If these
galaxies are confirmed to be starburst galaxies at $0.42 \lesssim z
\lesssim 0.84$, they would represent an intermediate stage between the
high redshift starburst population (i.e., Lyman-break galaxies), which
exhibit a characteristic rest frame optical luminosity $L^*_{LBG}
\simeq 6$-$9 L^*$ (Shapley et al.\ 2001) and can reach $L \gtrsim
20L^*$, and local starbursts which are generally compact (e.g.,
Brinchmann et al.\ 2004) with typical luminosities $L \lesssim$
1-2$L^*$ (Meurer et al.\ 2006).  The connection between strong,
low-ionization absorption and outflows from starburst galaxies has
already been established.  Schwartz et al.\ (2006), for example, find
\ion{C}{2} absorption with 1 \AA\ $ \lesssim W_0^{\lambda1335}
\lesssim 5.6$ \AA\ and 350 km s$^{-1}$ $\lesssim$ FWHM $\lesssim$ 1400
km s$^{-1}$ in outflows from individual star clusters in nearby
UV-selected galaxies.  Pettini et al.\ (2002) find low-ion (e.g.,
\ion{C}{2} $\lambda1334$, \ion{Si}{2} $\lambda1526$,
\ion{Al}{2}$\lambda1670$, \ion{Fe}{2} $\lambda2344$) absorption with
REWs $\simeq$ 2.6\AA\ - 3.5\AA\ in the spectrum of the $z=2.7$
gravitationally lensed LBG MS 1512$-$cB58, and Shapley et al.\ (2003)
find stacked spectra of $z \sim 3$ LBGs to have average \ion{C}{2}
$\lambda1334$ and \ion{Si}{2} $\lambda1526$ absorption FWHM $=560\pm
150$ km s$^{-1}$.  It is noteworthy that both of our strongest fields
that {\it do not} show evidence for a source with $\varepsilon < 1$,
{\it do} contain two $L(z=z_{abs}) \gtrsim 4 L^*$ galaxies with $b \le
90$ kpc.  Furthermore, in each of the three fields that have two such
bright galaxies, the quasar sightlines are ``bracketed'' by the bright
galaxies, i.e., the sightlines are intermediate in RA and/or DEC to
the galaxies.  If these galaxies in close proximity to the sightlines
are undergoing massive starbursts with galactic outflows at the
absorption redshift, they could easily explain the huge  $\Delta
V_{REW}$ seen in absorption.

Although our sample is not large and presently lacks galaxy-redshift
information, it does suggest a general picture of the nature of
ultra-strong \ion{Mg}{2} absorbers.  Many such systems are similar to
less-strong \ion{Mg}{2} absorbers, and may represent the tail of the
kinematic distribution caused by the chance alignment of many
low-ionization absorbing clouds.  Some of these cases are due to the
sightline passing in close proximity to, or through, the inner regions
of relatively bright galaxies that, due to interactions and/or star
formation, are kinematically more complex than relatively quiescent
galaxies such as the Milky Way.  These categories are not separate,
but may loosely correlate with absorption strength.  Additionally, a
smaller fraction of ultra-strong systems may be due to the winds of
very luminous star-forming galaxies with $4L \lesssim L^* \lesssim
13L^* $.  These tend to be among the strongest systems.  This proposed
dual-cause scenario is suggested by the results presented in
Table~\ref{table:environment} and is consistent with the results shown
in Figure~\ref{gal_numbers}, including the weaker excess of $L \approx
3L^*$ galaxies compared to those with $L \approx$ 1-2 $L^*$ and  $L
\gtrsim 4 L^*$.  Of course, larger samples would be needed to
constrain any real correlations; follow-up spectroscopy would greatly
aid in understanding individual fields and confirming or repudiating
the general picture suggested above.

\section{Conclusions}
\label{section:conclusions}

We have presented optical images of the fields surrounding 15 quasars
that exhibit intermediate redshift strong (two systems with 2.7 \AA\
$\lesssim$ \W\ $\lesssim 3$ \AA) or ultra-strong (13 systems with \W\
$\gtrsim 3$ \AA) \ion{Mg}{2} absorption in their spectra at redshifts
$0.42 < z_{abs} < 0.84$.  All of the fields reveal at least one
relatively bright galaxy within an impact parameter of 40 kpc to the
quasar sightline.  We have demonstrated a statistical overdensity of
galaxies within $b \lesssim 90$ kpc of the absorber sightlines
compared to predictions for randomly selected fields, and have shown
that this excess extends over a large range of luminosity.  There is
evidence for an excess of galaxy pairs, interactions, and very bright
($4 L^* \lesssim L \lesssim 13 L^*$) galaxies close to the sightlines
as well.  A significant number of our fields appear similar to those
for less-strong systems, but many of the ultra-strong \ion{Mg}{2}
absorber sightlines pass through the apparent stellar luminous extent
of relatively bright galaxies.  Since these absorbers exhibit velocity
spreads much larger than those expected for sightlines passing through
the Milky Way, these observations indicate the existence of
intermediate redshift galaxies that are more kinematically complex
than the Milky Way and a method for identifying them.  Interacting
galaxy pairs and starburst activity may in part contribute to the very
large kinematic spreads that define these systems.  Finally, our
results provide evidence that galactic winds from a population of
bright galaxies, that are intermediate to high-redshift Lyman break
galaxies and local starbursts, may contribute to the ultra-strong
\ion{Mg}{2} absorber population.  It would be very helpful to obtain
spectra of individual galaxies in the fields discussed in this paper,
as this would eliminate many of the ambiguities in the interpretations
of the individual fields.

\acknowledgments  

We wish to thank the the NOAO WIYN staff as well as Michele
Belfort-Mihalyi for help in acquiring our data.  DBN acknowledges
support from NSF grant AST-9984040.  DAT and SMR acknowledge support
from NSF grant AST-0307743.  AMQ acknowledges REU support from NSF
grant AST-0307743.

We thank members of the SDSS collaboration who made the SDSS project
a success. Funding for creation and distribution of the SDSS Archive
has been provided by the Alfred P. Sloan Foundation, Participating
Institutions, NASA, NSF, DOE, the Japanese Monbukagakusho, and the
Max-Planck Society. The SDSS Web site is http://www.sdss.org. The
SDSS is managed by the Astrophysical Research Consortium for the
Participating Institutions: the American Museum of Natural History,
Astrophysical Institute Potsdam, University of Basel, Cambridge
University, Case Western Reserve University, University of Chicago,
Drexel University, Fermilab, the Institute for Advanced Study,
the Japan Participation Group, Johns Hopkins University, the Joint
Institute for Nuclear Astrophysics, the Kavli Institute for Particle
Astrophysics and Cosmology, the Korean Scientist Group, the Chinese
Academy of Sciences (LAMOST), Los Alamos National Laboratory, the
Max-Planck-Institute for Astronomy (MPIA), the Max-Planck-Institute
for Astrophysics (MPA), New Mexico State University, Ohio State
University, University of Pittsburgh, University of Portsmouth,
Princeton University, the United States Naval Observatory, and the
University of Washington

\begin{deluxetable}{ccccccc}
\startdata
\hline\hline\\
\multicolumn{7}{c}{Q0013+1414, $z_{abs}=0.4838$, $W_0^{\lambda 2796}=2.69$ \AA}\\
[1ex]
$\Delta \alpha$ & $\Delta \delta$ &   &  $M_r$ at  & $L/L^*$ at   & $b$  & projected \\
(arcsec) & (arcsec) & $m_r$ & $z=0.4838$\tablenotemark{a} & $z=0.4838$\tablenotemark{a} & (kpc) & size\tablenotemark{b}\ (kpc) \\
\hline\\
1.5 & $-$2.1 & 22.00 & $-$20.49/$-$20.76/$-$20.90 & 0.52/0.66/0.75 & 15.5 & $17 \times 14$ \\
5.8 & 4.3 & 19.68 & $-$22.82/$-$23.08/$-$23.23 & 4.40/5.62/6.40 & 43.6 & $50 \times 35$ \\
$-$9.4 & 16.3 & 21.31 & $-$21.19/$-$21.45/$-$21.59 & 0.98/1.25/1.42 & 112.9 & $23 \times 18$ \\
$-$7.1 & 27.1 & 21.69 & $-$20.81/$-$21.08/$-$21.22 & 0.69/0.88/1.01 & 167.9 & $17 \times 14$ \\
29.4 & 1.5 & 22.95 & $-$19.55/$-$19.81/$-$19.95 & 0.22/0.28/0.31 & 176.7 & $10 \times  9$ \\
[1ex]
\hline\hline\\
\multicolumn{7}{c}{Q0232$-$0811, $z_{abs}=0.4523$, $W_0^{\lambda 2796}=3.69$ \AA}\\
[1ex]
$\Delta \alpha$ & $\Delta \delta$ &   &  $M_r$ at  & $L/L^*$ at   & $b$  & projected \\
(arcsec) & (arcsec) & $m_r$ & $z=0.4523$\tablenotemark{a} & $z=0.4523$\tablenotemark{a} & (kpc) & size\tablenotemark{b}\ (kpc) \\
\hline\\
0.0 & 0.8 & 22.0\tablenotemark{c} & $-$20.3/$-$20.5/$-$20.6 & 0.4/0.5/0.6 & 4.8 & $18 \times 12$ \\
3.0 & $-$8.6 & 21.35 & $-$20.93/$-$21.17/$-$21.29 & 0.77/0.96/1.08 & 52.6 & $33 \times 16$ \\
0.2 & $-$16.6 & 22.61 & $-$19.68/$-$19.91/$-$20.04 & 0.24/0.30/0.34 & 96.0 & $12 \times 10$ \\
17.0 & 13.9 & 20.09 & $-$22.19/$-$22.42/$-$22.55 & 2.47/3.06/3.44 & 127.0 & $30 \times 28$ \\
$-$4.1 & 23.3 & 22.76 & $-$19.52/$-$19.76/$-$19.88 & 0.21/0.26/0.29 & 136.9 & $21 \times  9$ \\
23.2 & 9.6 & 23.96 & $-$18.33/$-$18.56/$-$18.69 & 0.07/0.09/0.10 & 144.8 & $ 6 \times  6$ \\
26.2 & 4.0 & 24.54 & $-$17.74/$-$17.98/$-$18.10 & 0.04/0.05/0.06 & 152.9 & $ 4 \times  3$ \\
17.4 & $-$21.7 & 23.82 & $-$18.46/$-$18.70/$-$18.82 & 0.08/0.10/0.11 & 160.6 & $ 9 \times  8$ \\
19.7 & 21.7 & 22.65 & $-$19.63/$-$19.87/$-$20.00 & 0.23/0.29/0.33 & 169.4 & $17 \times 16$ \\
13.1 & $-$26.5 & 21.56 & $-$20.72/$-$20.96/$-$21.08 & 0.64/0.79/0.89 & 170.5 & $25 \times 19$ \\
$-$26.1 & $-$15.3 & 22.29 & $-$19.99/$-$20.23/$-$20.35 & 0.33/0.40/0.45 & 174.9 & $15 \times 12$ \\
$-$22.8 & $-$20.6 & 22.79 & $-$19.50/$-$19.73/$-$19.86 & 0.21/0.26/0.29 & 177.7 & $16 \times  9$ \\
28.2 & $-$12.6 & 24.20 & $-$18.08/$-$18.32/$-$18.44 & 0.06/0.07/0.08 & 178.3 & $ 8 \times  4$ \\
13.6 & $-$31.5 & 24.55 & $-$17.73/$-$17.97/$-$18.09 & 0.04/0.05/0.06 & 198.3 & $ 7 \times  5$ \\
[1ex]
\hline\hline\\
\multicolumn{7}{c}{Q0240$-$0812, $z_{abs}=0.5314$, $W_0^{\lambda 2796}=2.91$ \AA }\\
[1ex]
$\Delta \alpha$ & $\Delta \delta$ &   &  $M_r$ at  & $L/L^*$ at   & $b$  & projected \\
(arcsec) & (arcsec) & $m_r$ & $z=0.5314$\tablenotemark{a} & $z=0.5314$\tablenotemark{a} & (kpc) & size\tablenotemark{b}\ (kpc) \\
\hline\\
2.8 & 0.9 & 20.96 & $-$21.84/$-$22.15/$-$22.32 & 1.78/2.38/2.77 & 18.6 & $19 \times 17$ \\
$-$8.7 & 8.5 & 22.68 & $-$20.11/$-$20.42/$-$20.59 & 0.36/0.48/0.56 & 76.4 & $12 \times  6$ \\
$-$4.7 & 11.2 & 21.27 & $-$21.52/$-$21.84/$-$22.00 & 1.33/1.78/2.08 & 76.7 & $16 \times 15$ \\
13.3 & $-$2.0 & 22.75 & $-$20.04/$-$20.36/$-$20.52 & 0.34/0.46/0.53 & 85.0 & $ 9 \times  4$ \\
4.9 & $-$12.7 & 21.80 & $-$20.99/$-$21.30/$-$21.47 & 0.82/1.09/1.27 & 85.9 & $19 \times 11$ \\
$-$3.5 & $-$19.6 & 22.72 & $-$20.07/$-$20.39/$-$20.55 & 0.35/0.47/0.55 & 125.4 & $10 \times  7$ \\
8.2 & 26.5 & 21.25 & $-$21.55/$-$21.86/$-$22.03 & 1.36/1.82/2.12 & 174.9 & $16 \times 13$ \\
$-$28.2 & $-$6.8 & 21.49 & $-$21.30/$-$21.62/$-$21.78 & 1.09/1.45/1.69 & 182.8 & $17 \times 14$ \\
27.8 & 8.4 & 23.59 & $-$19.20/$-$19.51/$-$19.68 & 0.16/0.21/0.24 & 182.9 & $ 7 \times  5$ \\
$-$8.4 & 29.5 & 22.84 & $-$19.95/$-$20.27/$-$20.43 & 0.31/0.42/0.49 & 193.0 & $ 9 \times  8$ \\
[1ex]
\hline\hline\\
\multicolumn{7}{c}{Q0747+3054, $z_{abs}=0.7650$, $W_0^{\lambda 2796}=3.63$ \AA}\\
[1ex]
$\Delta \alpha$ & $\Delta \delta$ &   &  $M_i$ at  & $L/L^*$ at   & $b$  & projected \\
(arcsec) & (arcsec) & $m_i$ & $z=0.7650$\tablenotemark{a} & $z=0.7650$\tablenotemark{a} & (kpc) & size\tablenotemark{b}\ (kpc) \\
\hline\\
$-$1.9 & 4.7 & 19.53 & $-$24.20/$-$24.50/$-$24.70 & 11.1/14.6/17.6 & 37.7 & $47 \times 21$ \\
4.1 & 3.1 & 21.86 & $-$21.87/$-$22.17/$-$22.37 & 1.29/1.70/2.05 & 37.9 & $27 \times 16$ \\
$-$5.4 & $-$2.9 & 23.83 & $-$19.90/$-$20.20/$-$20.40 & 0.21/0.28/0.33 & 45.0 & $ 8 \times  8$ \\
$-$2.4 & $-$6.0 & 24.33 & $-$19.40/$-$19.70/$-$19.90 & 0.13/0.18/0.21 & 47.9 & $ 9 \times  8$ \\
$-$7.8 & $-$1.8 & 22.99 & $-$20.73/$-$21.04/$-$21.24 & 0.45/0.60/0.72 & 59.1 & $16 \times 11$ \\
6.8 & 5.3 & 20.67 & $-$23.06/$-$23.37/$-$23.57 & 3.88/5.13/6.17 & 63.4 & $35 \times 26$ \\
$-$8.4 & $-$5.9 & 24.10 & $-$19.63/$-$19.94/$-$20.14 & 0.16/0.22/0.26 & 75.8 & $11 \times  6$ \\
6.5 & $-$9.9 & 22.68 & $-$21.05/$-$21.36/$-$21.56 & 0.61/0.81/0.97 & 87.7 & $18 \times 13$ \\
$-$11.0 & 5.0 & 24.30 & $-$19.43/$-$19.73/$-$19.93 & 0.14/0.18/0.22 & 89.1 & $ 9 \times  5$ \\
$-$12.1 & $-$1.6 & 24.17 & $-$19.56/$-$19.86/$-$20.06 & 0.15/0.20/0.24 & 90.4 & $ 8 \times  7$ \\
11.3 & 8.1 & 23.19 & $-$20.54/$-$20.84/$-$21.04 & 0.38/0.50/0.61 & 102.9 & $13 \times 11$ \\
9.5 & 11.3 & 22.73 & $-$21.00/$-$21.31/$-$21.51 & 0.58/0.77/0.93 & 108.8 & $18 \times 13$ \\
$-$12.9 & 7.6 & 23.98 & $-$19.75/$-$20.05/$-$20.25 & 0.18/0.24/0.29 & 110.4 & $10 \times  7$ \\
$-$5.9 & $-$15.0 & 23.75 & $-$19.97/$-$20.28/$-$20.48 & 0.23/0.30/0.36 & 119.0 & $12 \times  6$ \\
$-$16.1 & 7.9 & 24.08 & $-$19.65/$-$19.95/$-$20.15 & 0.17/0.22/0.27 & 132.5 & $10 \times  8$ \\
$-$13.4 & 12.2 & 23.65 & $-$20.07/$-$20.38/$-$20.58 & 0.25/0.33/0.39 & 134.0 & $ 9 \times  8$ \\
$-$15.9 & 14.4 & 24.01 & $-$19.72/$-$20.02/$-$20.22 & 0.18/0.24/0.28 & 158.2 & $ 9 \times  8$ \\
$-$12.5 & 18.3 & 23.76 & $-$19.97/$-$20.28/$-$20.48 & 0.23/0.30/0.36 & 163.7 & $ 9 \times  8$ \\
$-$22.9 & $-$3.2 & 23.91 & $-$19.82/$-$20.13/$-$20.33 & 0.20/0.26/0.31 & 171.0 & $ 8 \times  6$ \\
13.3 & $-$20.0 & 24.80 & $-$18.93/$-$19.23/$-$19.43 & 0.09/0.11/0.14 & 177.7 & $ 9 \times  4$ \\
17.6 & 17.3 & 23.99 & $-$19.73/$-$20.04/$-$20.24 & 0.18/0.24/0.29 & 182.4 & $ 9 \times  8$ \\
16.0 & $-$19.3 & 23.86 & $-$19.87/$-$20.17/$-$20.37 & 0.20/0.27/0.33 & 185.4 & $ 9 \times  8$ \\
$-$23.1 & $-$9.9 & 22.72 & $-$21.01/$-$21.31/$-$21.51 & 0.59/0.78/0.93 & 185.8 & $17 \times 13$ \\
16.9 & 19.3 & 22.79 & $-$20.94/$-$21.24/$-$21.44 & 0.55/0.73/0.87 & 189.4 & $12 \times 12$ \\
8.7 & 24.3 & 24.84 & $-$18.89/$-$19.20/$-$19.40 & 0.08/0.11/0.13 & 190.9 & $ 7 \times  6$ \\
$-$23.5 & $-$11.2 & 23.26 & $-$20.47/$-$20.78/$-$20.98 & 0.36/0.47/0.57 & 192.0 & $13 \times 11$ \\
[1ex]
\hline\hline\\
\multicolumn{7}{c}{Q0747+3354, $z_{abs}=0.6202$, $W_0^{\lambda 2796}=4.69$ \AA}\\
[1ex]
$\Delta \alpha$ & $\Delta \delta$ &   &  $M_i$ at  & $L/L^*$ at   & $b$  & projected \\
(arcsec) & (arcsec) & $m_i$ & $z=0.6202$\tablenotemark{a} & $z=0.6202$\tablenotemark{a} & (kpc) & size\tablenotemark{b}\ (kpc) \\
\hline\\
1.7 & 0.9 & 21.11 & $-$21.95/$-$22.16/$-$22.32 & 1.39/1.70/1.96 & 12.8 & $32 \times 22$ \\
0.2 & 5.1 & 23.14 & $-$19.92/$-$20.14/$-$20.29 & 0.22/0.26/0.30 & 34.3 & $11 \times  9$ \\
11.4 & $-$10.3 & 22.20 & $-$20.87/$-$21.08/$-$21.24 & 0.51/0.62/0.72 & 104.4 & $16 \times 15$ \\
$-$18.0 & $-$5.0 & 22.55 & $-$20.52/$-$20.73/$-$20.89 & 0.37/0.45/0.52 & 126.6 & $18 \times 14$ \\
$-$11.7 & $-$14.7 & 23.72 & $-$19.35/$-$19.56/$-$19.72 & 0.13/0.15/0.18 & 127.7 & $ 9 \times  4$ \\
$-$19.0 & $-$6.5 & 21.86 & $-$21.21/$-$21.42/$-$21.58 & 0.70/0.86/0.99 & 136.5 & $18 \times 17$ \\
$-$2.9 & 25.5 & 21.69 & $-$21.38/$-$21.59/$-$21.75 & 0.82/1.00/1.15 & 174.3 & $18 \times 17$ \\
[1ex]
\hline\hline\\
\multicolumn{7}{c}{Q0800+2150, $z_{abs}=0.5716$, $W_0^{\lambda 2796}=3.65$ \AA}\\
[1ex]
$\Delta \alpha$ & $\Delta \delta$ &   &  $M_r$ at  & $L/L^*$ at   & $b$  & projected \\
(arcsec) & (arcsec) & $m_r$ & $z=0.5716$\tablenotemark{a} & $z=0.5716$\tablenotemark{a} & (kpc) & size\tablenotemark{b}\ (kpc) \\
\hline\\
2.6 & 1.2 & 22.22 & $-$20.84/$-$21.20/$-$21.39 & 0.71/0.99/1.19 & 18.6 & $21 \times 15$ \\
2.6 & $-$1.3 & 24.63 & $-$18.42/$-$18.79/$-$18.98 & 0.08/0.11/0.13 & 19.0 & $ 9 \times  8$ \\
$-$0.8 & $-$6.7 & 23.98 & $-$19.08/$-$19.44/$-$19.63 & 0.14/0.20/0.23 & 44.1 & $11 \times  8$ \\
6.2 & 5.9 & 24.73 & $-$18.33/$-$18.69/$-$18.88 & 0.07/0.10/0.12 & 56.0 & $12 \times  4$ \\
$-$8.3 & 2.5 & 23.66 & $-$19.39/$-$19.76/$-$19.95 & 0.19/0.26/0.31 & 56.7 & $12 \times 11$ \\
8.5 & 3.1 & 24.38 & $-$18.67/$-$19.04/$-$19.23 & 0.10/0.14/0.16 & 59.4 & $10 \times  9$ \\
7.4 & 13.7 & 23.28 & $-$19.78/$-$20.14/$-$20.33 & 0.27/0.37/0.45 & 102.1 & $16 \times 14$ \\
$-$7.3 & 14.1 & 22.28 & $-$20.78/$-$21.14/$-$21.33 & 0.67/0.94/1.12 & 103.4 & $20 \times 15$ \\
11.8 & $-$10.8 & 24.55 & $-$18.51/$-$18.87/$-$19.06 & 0.08/0.12/0.14 & 104.7 & $ 8 \times  7$ \\
6.3 & $-$15.7 & 24.96 & $-$18.10/$-$18.46/$-$18.65 & 0.06/0.08/0.10 & 110.4 & $ 7 \times  5$ \\
$-$17.7 & 4.7 & 21.96 & $-$21.10/$-$21.46/$-$21.65 & 0.90/1.26/1.50 & 119.6 & $12 \times 12$ \\
$-$2.8 & $-$23.5 & 24.89 & $-$18.16/$-$18.53/$-$18.72 & 0.06/0.08/0.10 & 154.4 & $ 8 \times  6$ \\
$-$9.6 & $-$22.8 & 23.15 & $-$19.90/$-$20.27/$-$20.46 & 0.30/0.42/0.50 & 161.7 & $15 \times 13$ \\
$-$23.3 & $-$9.7 & 25.00 & $-$18.06/$-$18.42/$-$18.61 & 0.05/0.08/0.09 & 165.2 & $ 7 \times  5$ \\
9.3 & $-$23.5 & 24.59 & $-$18.46/$-$18.83/$-$19.02 & 0.08/0.11/0.13 & 165.2 & $ 9 \times  6$ \\
$-$9.2 & 23.6 & 21.67 & $-$21.38/$-$21.75/$-$21.94 & 1.17/1.64/1.96 & 165.6 & $20 \times 14$ \\
9.8 & 25.1 & 24.38 & $-$18.67/$-$19.03/$-$19.23 & 0.10/0.13/0.16 & 175.8 & $ 9 \times  9$ \\
$-$6.5 & 26.6 & 24.85 & $-$18.20/$-$18.57/$-$18.76 & 0.06/0.09/0.10 & 179.0 & $ 6 \times  6$ \\
14.7 & $-$23.8 & 24.68 & $-$18.38/$-$18.74/$-$18.93 & 0.07/0.10/0.12 & 182.7 & $ 9 \times  7$ \\
$-$18.2 & 21.2 & 24.22 & $-$18.83/$-$19.20/$-$19.39 & 0.11/0.16/0.19 & 182.7 & $10 \times  7$ \\
25.9 & $-$12.0 & 19.81 & $-$23.25/$-$23.61/$-$23.80 & 6.52/9.12/10.9 & 186.7 & $60 \times 38$ \\
11.6 & $-$26.8 & 23.85 & $-$19.21/$-$19.57/$-$19.76 & 0.16/0.22/0.26 & 190.9 & $14 \times  8$ \\
$-$21.8 & $-$19.5 & 23.24 & $-$19.81/$-$20.18/$-$20.37 & 0.28/0.39/0.46 & 191.3 & $12 \times 11$ \\
$-$0.2 & $-$29.4 & 21.59 & $-$21.47/$-$21.83/$-$22.02 & 1.27/1.77/2.11 & 192.3 & $20 \times 15$ \\
13.5 & $-$26.6 & 23.46 & $-$19.60/$-$19.96/$-$20.15 & 0.23/0.32/0.38 & 194.9 & $13 \times 11$ \\
[1ex]
\hline\hline\\
\multicolumn{7}{c}{Q0836+5132, $z_{abs}=0.5666$, $W_0^{\lambda 2796}=3.48$ \AA}\\
[1ex]
$\Delta \alpha$ & $\Delta \delta$ &   &  $M_r$ at  & $L/L^*$ at   & $b$  & projected \\
(arcsec) & (arcsec) & $m_r$ & $z=0.5666$\tablenotemark{a} & $z=0.5666$\tablenotemark{a} & (kpc) & size\tablenotemark{b}\ (kpc) \\
\hline\\
$-$0.6 & $-$1.4 & 23.02 & $-$20.01/$-$20.36/$-$20.55 & 0.33/0.46/0.55 & 10.1 & $22 \times 18$ \\
9.2 & $-$1.4 & 24.74 & $-$18.29/$-$18.65/$-$18.83 & 0.07/0.09/0.11 & 60.4 & $ 6 \times  4$ \\
3.3 & $-$11.6 & 23.18 & $-$19.84/$-$20.20/$-$20.38 & 0.28/0.39/0.47 & 78.6 & $18 \times 16$ \\
4.1 & 17.0 & 21.68 & $-$21.34/$-$21.70/$-$21.89 & 1.13/1.57/1.87 & 113.9 & $35 \times 26$ \\
5.8 & 19.6 & 20.71 & $-$22.32/$-$22.68/$-$22.86 & 2.78/3.86/4.59 & 133.1 & $45 \times 38$ \\
$-$19.7 & $-$10.2 & 23.51 & $-$19.51/$-$19.87/$-$20.06 & 0.21/0.29/0.35 & 144.0 & $11 \times  8$ \\
$-$16.1 & $-$20.9 & 17.09\tablenotemark{d} & \nodata & \nodata & 172.0 & \nodata \\
16.4 & 21.8 & 20.84 & $-$22.18/$-$22.54/$-$22.73 & 2.45/3.41/4.05 & 177.9 & $52 \times 31$ \\
[1ex]
\hline\hline\\
\multicolumn{7}{c}{Q0902+3722, $z_{abs}=0.6697$, $W_0^{\lambda 2796}=3.97$ \AA }\\
[1ex]
$\Delta \alpha$ & $\Delta \delta$ &   &  $M_i$ at  & $L/L^*$ at   & $b$  & projected \\
(arcsec) & (arcsec) & $m_i$ & $z=0.6697$\tablenotemark{a} & $z=0.6697$\tablenotemark{a} & (kpc) & size\tablenotemark{b}\ (kpc) \\
\hline\\
2.2 & $-$1.4 & 21.42 & $-$21.89/$-$22.13/$-$22.30 & 1.32/1.64/1.92 & 18.4 & $20 \times 18$ \\
$-$4.6 & 2.3 & 22.38 & $-$20.92/$-$21.17/$-$21.34 & 0.54/0.68/0.79 & 35.8 & $16 \times 14$ \\
4.8 & $-$2.1 & 21.70 & $-$21.61/$-$21.85/$-$22.02 & 1.02/1.27/1.49 & 36.8 & $29 \times 15$ \\
$-$5.4 & $-$12.1 & 22.60 & $-$20.71/$-$20.95/$-$21.12 & 0.44/0.55/0.65 & 92.8 & $11 \times 11$ \\
$-$10.7 & 9.4 & 23.93 & $-$19.37/$-$19.61/$-$19.78 & 0.13/0.16/0.19 & 99.8 & $ 9 \times  5$ \\
14.5 & 10.0 & 22.99 & $-$20.31/$-$20.55/$-$20.73 & 0.31/0.39/0.45 & 123.4 & $10 \times  6$ \\
$-$0.0 & $-$18.7 & 23.13 & $-$20.18/$-$20.42/$-$20.59 & 0.27/0.34/0.40 & 131.2 & $ 9 \times  8$ \\
$-$7.8 & 17.8 & 21.69 & $-$21.61/$-$21.85/$-$22.02 & 1.02/1.27/1.49 & 136.5 & $24 \times 16$ \\
$-$8.6 & $-$19.8 & 23.94 & $-$19.37/$-$19.61/$-$19.78 & 0.13/0.16/0.19 & 151.6 & $ 8 \times  5$ \\
4.6 & 21.5 & 23.38 & $-$19.92/$-$20.16/$-$20.33 & 0.22/0.27/0.31 & 154.1 & $10 \times  9$ \\
6.9 & 23.7 & 20.30 & $-$23.00/$-$23.24/$-$23.41 & 3.67/4.58/5.36 & 172.8 & $24 \times 16$ \\
$-$24.8 & 4.6 & 20.69 & $-$22.62/$-$22.86/$-$23.03 & 2.58/3.22/3.77 & 177.0 & $42 \times 17$ \\
$-$24.7 & 8.9 & 21.07 & $-$22.24/$-$22.48/$-$22.65 & 1.82/2.27/2.65 & 184.4 & $34 \times 15$ \\
$-$11.2 & $-$23.9 & 23.45 & $-$19.85/$-$20.09/$-$20.26 & 0.20/0.25/0.29 & 185.4 & $ 9 \times  8$ \\
$-$3.3 & 26.9 & 23.21 & $-$20.09/$-$20.33/$-$20.51 & 0.25/0.31/0.37 & 190.3 & $10 \times  9$ \\
[1ex]
\hline\hline\\
\multicolumn{7}{c}{Q1000+4438, $z_{abs}=0.7192$, $W_0^{\lambda 2796}=5.33$ \AA}\\
[1ex]
$\Delta \alpha$ & $\Delta \delta$ &   &  $M_i$ at  & $L/L^*$ at   & $b$  & projected \\
(arcsec) & (arcsec) & $m_i$ & $z=0.7192$\tablenotemark{a} & $z=0.7192$\tablenotemark{a} & (kpc) & size\tablenotemark{b}\ (kpc) \\
\hline\\
0.8 & 0.8 & 21.5\tablenotemark{c} & $-$22.0/$-$22.3/$-$22.5 & 1.5/1.9/2.3 & 8.4 & $28 \times 22$ \\
4.1 & 4.6 & 23.94 & $-$19.60/$-$19.87/$-$20.05 & 0.16/0.20/0.24 & 44.4 & $11 \times 10$ \\
5.7 & 13.7 & 22.77 & $-$20.76/$-$21.04/$-$21.22 & 0.47/0.60/0.71 & 107.2 & $18 \times 11$ \\
$-$17.2 & $-$5.8 & 20.91 & $-$22.62/$-$22.89/$-$23.08 & 2.58/3.32/3.94 & 131.1 & $33 \times 32$ \\
1.8 & 19.8 & 22.73 & $-$20.80/$-$21.07/$-$21.26 & 0.48/0.62/0.74 & 144.0 & $17 \times  9$ \\
19.4 & 7.3 & 23.06 & $-$20.48/$-$20.75/$-$20.94 & 0.36/0.46/0.55 & 149.7 & $18 \times 16$ \\
4.8 & 22.4 & 21.31 & $-$22.22/$-$22.49/$-$22.68 & 1.79/2.30/2.72 & 165.5 & $31 \times 22$ \\
$-$18.7 & $-$18.1 & 22.57 & $-$20.97/$-$21.24/$-$21.42 & 0.56/0.72/0.86 & 188.1 & $19 \times 16$ \\
$-$17.2 & $-$19.6 & 21.22 & $-$22.32/$-$22.59/$-$22.77 & 1.95/2.51/2.98 & 188.7 & $22 \times 20$ \\
13.0 & $-$24.3 & 23.85 & $-$19.69/$-$19.96/$-$20.14 & 0.17/0.22/0.26 & 199.4 & $ 8 \times  6$ \\
$-$27.5 & $-$2.6 & 23.80 & $-$19.73/$-$20.00/$-$20.19 & 0.18/0.23/0.27 & 199.7 & $11 \times  9$ \\
[1ex]
\hline\hline\\
\multicolumn{7}{c}{Q1011+4451, $z_{abs}=0.8360$, $W_0^{\lambda 2796}=4.94$ \AA}\\
[1ex]
$\Delta \alpha$ & $\Delta \delta$ &   &  $M_i$ at  & $L/L^*$ at   & $b$  & projected \\
(arcsec) & (arcsec) & $m_i$ & $z=0.8360$\tablenotemark{a} & $z=0.8360$\tablenotemark{a} & (kpc) & size\tablenotemark{b}\ (kpc) \\
\hline\\
$-$0.1 & 2.6 & 24.45 & $-$19.59/$-$19.95/$-$20.18 & 0.16/0.22/0.27 & 20.0 & $ 8 \times 6$ \\
1.5 & $-$3.9 & 20.63 & $-$23.41/$-$23.77/$-$24.00 & 5.34/7.44/9.18 & 32.1 & $27 \times 22$ \\
$-$5.8 & 1.0 & 21.10 & $-$22.95/$-$23.31/$-$23.53 & 3.49/4.86/5.99 & 44.8 & $29 \times 23$ \\
$-$4.3 & 4.3 & 24.10 & $-$19.94/$-$20.30/$-$20.53 & 0.22/0.31/0.38 & 46.3 & $13 \times 5$ \\
7.9 & 5.5 & 23.79 & $-$20.25/$-$20.61/$-$20.84 & 0.29/0.41/0.50 & 73.6 & $11 \times 9$ \\
8.7 & $-$5.7 & 23.35 & $-$20.69/$-$21.05/$-$21.28 & 0.44/0.61/0.75 & 79.5 & $12 \times 11$ \\
$-$8.7 & 8.1 & 24.58 & $-$19.46/$-$19.82/$-$20.05 & 0.14/0.20/0.24 & 90.7 & $ 9 \times 6$ \\
9.2 & $-$10.0 & 23.79 & $-$20.25/$-$20.61/$-$20.84 & 0.29/0.41/0.50 & 103.3 & $15 \times 7$ \\
12.9 & 11.0 & 24.74 & $-$19.30/$-$19.66/$-$19.89 & 0.12/0.17/0.21 & 128.9 & $10 \times 5$ \\
9.6 & $-$16.9 & 22.87 & $-$21.17/$-$21.53/$-$21.76 & 0.68/0.95/1.17 & 148.0 & $15 \times 11$ \\
$-$16.7 & 11.2 & 21.78 & $-$22.27/$-$22.63/$-$22.85 & 1.86/2.60/3.20 & 153.1 & $20 \times 17$ \\
19.9 & 3.6 & 23.82 & $-$20.22/$-$20.58/$-$20.81 & 0.28/0.39/0.49 & 153.9 & $16 \times 8$ \\
$-$10.2 & $-$18.9 & 24.10 & $-$19.94/$-$20.30/$-$20.53 & 0.22/0.31/0.38 & 163.6 & $12 \times 9$ \\
20.2 & 7.6 & 22.66 & $-$21.38/$-$21.74/$-$21.97 & 0.82/1.15/1.41 & 164.2 & $18 \times 16$ \\
$-$8.0 & 20.4 & 24.89 & $-$19.15/$-$19.51/$-$19.74 & 0.11/0.15/0.18 & 166.8 & $ 8 \times 5$ \\
$-$19.9 & $-$9.3 & 21.62 & $-$22.42/$-$22.78/$-$23.01 & 2.15/2.99/3.69 & 167.2 & $16 \times 15$ \\
20.4 & 10.1 & 23.45 & $-$20.59/$-$20.95/$-$21.18 & 0.40/0.56/0.69 & 173.6 & $11 \times 10$ \\
$-$24.4 & $-$0.3 & 22.54 & $-$21.51/$-$21.87/$-$22.09 & 0.92/1.29/1.59 & 185.8 & $14 \times 12$ \\
$-$24.4 & $-$8.4 & 16.78 & $-$27.26/$-$27.62/$-$27.85 & 186/259/320 & 196.8 & $192 \times 35$ \\
[1ex]
\hline\hline\\
\multicolumn{7}{c}{Q1038+4727, $z_{abs}=0.5292$, $W_0^{\lambda 2796}=3.14$ \AA}\\
[1ex]
$\Delta \alpha$ & $\Delta \delta$ &   &  $M_r$ at  & $L/L^*$ at   & $b$  & projected \\
(arcsec) & (arcsec) & $m_r$ & $z=0.5292$\tablenotemark{a} & $z=0.5292$\tablenotemark{a} & (kpc) & size\tablenotemark{b}\ (kpc) \\
\hline\\
$-$1.9 & 1.3 & 21.12 & $-$21.66/$-$21.97/$-$22.13 & 1.51/2.01/2.34 & 14.8 & $28 \times 16$ \\
$-$4.6 & 3.0 & 23.01 & $-$19.78/$-$20.09/$-$20.25 & 0.27/0.36/0.41 & 34.9 & $12 \times  7$ \\
$-$15.9 & 3.1 & 21.33 & $-$21.46/$-$21.77/$-$21.93 & 1.25/1.67/1.95 & 102.1 & $22 \times 20$ \\
7.1 & 21.7 & 22.64 & $-$20.14/$-$20.45/$-$20.62 & 0.37/0.50/0.58 & 143.8 & $15 \times 12$ \\
$-$22.9 & 8.6 & 22.75 & $-$20.03/$-$20.35/$-$20.51 & 0.34/0.45/0.52 & 153.8 & $12 \times 11$ \\
$-$24.4 & 4.7 & 22.78 & $-$20.00/$-$20.31/$-$20.47 & 0.33/0.44/0.51 & 156.5 & $11 \times 10$ \\
21.6 & 13.3 & 23.28 & $-$19.50/$-$19.82/$-$19.98 & 0.21/0.28/0.32 & 159.6 & $10 \times  9$ \\
21.8 & $-$19.2 & 23.43 & $-$19.36/$-$19.67/$-$19.83 & 0.18/0.24/0.28 & 182.8 & $14 \times  7$ \\
27.5 & $-$13.9 & 23.57 & $-$19.21/$-$19.53/$-$19.69 & 0.16/0.21/0.25 & 193.7 & $10 \times  6$ \\
1.5 & 30.8 & 22.38 & $-$20.40/$-$20.71/$-$20.87 & 0.47/0.63/0.73 & 194.0 & $13 \times 12$ \\
[1ex]
\hline\hline\\
\multicolumn{7}{c}{Q1356+6119, $z_{abs}=0.7850$, $W_0^{\lambda 2796}=5.97$ \AA}\\
[1ex]
$\Delta \alpha$ & $\Delta \delta$ &   &  $M_i$ at  & $L/L^*$ at   & $b$  & projected \\
(arcsec) & (arcsec) & $m_i$ & $z=0.7850$\tablenotemark{a} & $z=0.7850$\tablenotemark{a} & (kpc) & size\tablenotemark{b}\ (kpc) \\
\hline\\
$-$1.0 & 0.5 & 21.7\tablenotemark{c} & $-$22.1/$-$22.5/$-$22.7 & 1.7/2.2/2.7 & 8.6 & $22 \times 17$ \\
$-$0.3 & 6.7 & 22.02 & $-$21.79/$-$22.11/$-$22.32 & 1.21/1.62/1.96 & 50.4 & $21 \times 17$ \\
7.3 & 6.5 & 22.81 & $-$21.01/$-$21.33/$-$21.53 & 0.59/0.79/0.95 & 72.7 & $18 \times 14$ \\
$-$1.5 & 10.6 & 23.15 & $-$20.66/$-$20.98/$-$21.19 & 0.43/0.57/0.69 & 79.5 & $15 \times 12$ \\
13.3 & $-$7.8 & 22.05\tablenotemark{e} & $-$21.77/$-$22.09/$-$22.29 & 1.18/1.58/1.91 & 114.8 & $16 \times 14$ \\
14.3 & $-$9.0 & 24.05 & $-$19.76/$-$20.08/$-$20.29 & 0.19/0.25/0.30 & 126.2 & $12 \times  6$ \\
$-$7.6 & 17.7 & 21.95 & $-$21.87/$-$22.19/$-$22.39 & 1.29/1.73/2.10 & 143.7 & $17 \times 15$ \\
18.8 & 6.6 & 23.90 & $-$19.91/$-$20.23/$-$20.44 & 0.21/0.29/0.35 & 148.8 & $ 8 \times  7$ \\
$-$1.8 & 21.4 & 19.94 & $-$23.87/$-$24.19/$-$24.40 & 8.19/11.0/13.3 & 159.8 & $37 \times 35$ \\
19.7 & $-$8.7 & 22.41 & $-$21.41/$-$21.73/$-$21.93 & 0.85/1.13/1.37 & 161.0 & $19 \times 15$ \\
$-$21.6 & $-$5.2 & 20.32 & $-$23.49/$-$23.81/$-$24.02 & 5.78/7.73/9.36 & 165.9 & $49 \times 25$ \\
20.9 & $-$8.2 & 23.36 & $-$20.46/$-$20.78/$-$20.98 & 0.35/0.47/0.57 & 167.5 & $14 \times 10$ \\
$-$15.6 & $-$19.8 & 20.74 & $-$23.08/$-$23.40/$-$23.60 & 3.94/5.28/6.39 & 188.0 & $28 \times 23$ \\
$-$24.8 & 5.3 & 19.64 & $-$24.18/$-$24.49/$-$24.70 & 10.8/14.5/17.6 & 188.9 & $39 \times 29$ \\
$-$17.8 & 19.0 & 18.81 & $-$25.01/$-$25.33/$-$25.53 & 23.3/31.2/37.8 & 194.0 & $75 \times 48$ \\
$-$7.4 & 25.0 & 23.14 & $-$20.68/$-$20.99/$-$21.20 & 0.43/0.58/0.70 & 194.6 & $14 \times 12$ \\
4.9 & 26.0 & 22.98 & $-$20.84/$-$21.15/$-$21.36 & 0.50/0.67/0.81 & 197.1 & $14 \times 10$ \\
$-$25.4 & 7.5 & 22.29 & $-$21.52/$-$21.84/$-$22.05 & 0.94/1.26/1.52 & 197.5 & $20 \times 16$ \\
$-$19.7 & $-$18.0 & 22.46 & $-$21.36/$-$21.68/$-$21.88 & 0.81/1.08/1.31 & 199.1 & $25 \times 15$ \\
[1ex]
\hline\hline\\
\multicolumn{7}{c}{Q1417+0115, $z_{abs}=0.6687$, $W_0^{\lambda 2796}=5.6$ \AA}\\
[1ex]
$\Delta \alpha$ & $\Delta \delta$ &   &  $M_i$ at  & $L/L^*$ at   & $b$  & projected \\
(arcsec) & (arcsec) & $m_i$ & $z=0.6687$\tablenotemark{a} & $z=0.6687$\tablenotemark{a} & (kpc) & size\tablenotemark{b}\ (kpc) \\
\hline\\
$-$2.5 & $-$3.3 & 21.86 & $-$21.44/$-$21.68/$-$21.85 & 0.87/1.09/1.28 & 29.0 & $19 \times 12$ \\
$-$6.4 & $-$5.0 & 23.26 & $-$20.04/$-$20.28/$-$20.45 & 0.24/0.30/0.35 & 57.1 & $ 7 \times  4$ \\
6.7 & 5.3 & 20.33 & $-$22.97/$-$23.21/$-$23.38 & 3.57/4.46/5.21 & 60.0 & $27 \times 25$ \\
11.0 & $-$2.4 & 23.01 & $-$20.29/$-$20.54/$-$20.71 & 0.30/0.38/0.44 & 79.2 & $12 \times  7$ \\
7.0 & $-$10.1 & 19.32 & $-$23.98/$-$24.22/$-$24.39 & 9.02/11.3/13.2 & 86.3 & $42 \times 38$ \\
18.4 & $-$1.2 & 21.15 & $-$22.15/$-$22.39/$-$22.56 & 1.68/2.09/2.45 & 129.2 & $26 \times 20$ \\
$-$11.6 & 14.8 & 22.34 & $-$20.96/$-$21.20/$-$21.37 & 0.56/0.70/0.81 & 132.1 & $15 \times  8$ \\
$-$23.0 & 1.9 & 21.27 & $-$22.03/$-$22.27/$-$22.44 & 1.50/1.87/2.18 & 162.1 & $26 \times 19$ \\
$-$20.5 & $-$12.1 & 20.78 & $-$22.52/$-$22.76/$-$22.93 & 2.35/2.93/3.43 & 166.5 & $30 \times 23$ \\
$-$9.7 & $-$23.2 & 21.90 & $-$21.40/$-$21.64/$-$21.81 & 0.84/1.05/1.23 & 176.5 & $18 \times 16$ \\
$-$19.3 & $-$20.3 & 22.36 & $-$20.94/$-$21.18/$-$21.35 & 0.55/0.68/0.80 & 196.4 & $13 \times 12$ \\
[1ex]
\hline\hline\\
\multicolumn{7}{c}{Q1427+5325, $z_{abs}=0.5537$, $W_0^{\lambda 2796}=4.35$ \AA}\\
[1ex]
$\Delta \alpha$ & $\Delta \delta$ &   &  $M_r$ at  & $L/L^*$ at   & $b$  & projected \\
(arcsec) & (arcsec) & $m_r$ & $z=0.5537$\tablenotemark{a} & $z=0.5537$\tablenotemark{a} & (kpc) & size\tablenotemark{b}\ (kpc) \\
\hline\\
1.7 & 0.7 & 21.3\tablenotemark{c} & $-$21.7/$-$22.0/$-$22.2 & 1.6/2.1/2.5 & 11.7 & $31 \times 25$ \\
0.4 & $-$2.4 & 22.3\tablenotemark{c} & $-$20.6/$-$20.9/$-$21.1 & 0.6/0.8/0.9 & 15.4 & $25 \times 19$ \\
2.7 & 2.3 & 22.1\tablenotemark{c} & $-$20.9/$-$21.2/$-$21.4 & 0.7/1.0/1.2 & 23.1 & $21 \times 20$ \\
$-$5.3 & 1.4 & 23.78\tablenotemark{e} & $-$19.16/$-$19.50/$-$19.68 & 0.15/0.21/0.24 & 35.5 & $ 9 \times  7$ \\
6.4 & $-$1.1 & 21.82 & $-$21.12/$-$21.46/$-$21.64 & 0.92/1.26/1.49 & 41.8 & $18 \times 18$ \\
$-$12.1 & 4.5 & 23.31 & $-$19.63/$-$19.97/$-$20.15 & 0.23/0.32/0.38 & 83.2 & $13 \times  8$ \\
14.8 & $-$0.8 & 21.48 & $-$21.46/$-$21.80/$-$21.98 & 1.26/1.72/2.03 & 95.1 & $34 \times 20$ \\
$-$15.7 & $-$14.5 & 22.98 & $-$19.96/$-$20.30/$-$20.48 & 0.32/0.43/0.51 & 137.3 & $15 \times 12$ \\
9.0 & 20.0 & 24.21 & $-$18.73/$-$19.07/$-$19.25 & 0.10/0.14/0.16 & 141.3 & $11 \times  4$ \\
$-$25.0 & $-$12.6 & 23.46 & $-$19.48/$-$19.82/$-$20.00 & 0.20/0.28/0.33 & 180.0 & $10 \times  8$ \\
$-$20.9 & $-$21.7 & 21.12 & $-$21.82/$-$22.16/$-$22.34 & 1.75/2.39/2.82 & 193.5 & $36 \times 30$ \\
[1ex]
\hline\hline\\
\multicolumn{7}{c}{Q1520+6105, $z_{abs}=0.4235$, $W_0^{\lambda 2796}=4.24$ \AA}\\
[1ex]
$\Delta \alpha$ & $\Delta \delta$ &   &  $M_r$ at  & $L/L^*$ at   & $b$  & projected \\
(arcsec) & (arcsec) & $m_r$ & $z=0.4235$\tablenotemark{a} & $z=0.4235$\tablenotemark{a} & (kpc) & size\tablenotemark{b}\ (kpc) \\
\hline\\
$-$2.7 & 2.3 & 19.72 & $-$22.36/$-$22.57/$-$22.69 & 2.89/3.51/3.90 & 19.6 & $31 \times 19$ \\
13.1 & $-$6.6 & 22.90 & $-$19.18/$-$19.39/$-$19.51 & 0.15/0.19/0.21 & 81.4 & $ 9 \times  8$ \\
15.8 & 13.1 & 21.00 & $-$21.08/$-$21.30/$-$21.41 & 0.89/1.08/1.20 & 113.9 & $14 \times 12$ \\
12.1 & $-$17.7 & 22.83 & $-$19.26/$-$19.47/$-$19.59 & 0.17/0.20/0.22 & 119.4 & $ 8 \times  7$ \\
2.0 & $-$23.7 & 22.56 & $-$19.52/$-$19.73/$-$19.85 & 0.21/0.26/0.28 & 132.3 & $12 \times 10$ \\
$-$9.0 & $-$22.7 & 20.25 & $-$21.84/$-$22.05/$-$22.17 & 1.78/2.17/2.41 & 135.9 & $25 \times 24$ \\
20.1 & $-$15.7 & 19.91 & $-$22.18/$-$22.39/$-$22.50 & 2.43/2.96/3.29 & 142.0 & $23 \times 20$ \\
11.8 & $-$24.6 & 22.84 & $-$19.24/$-$19.45/$-$19.57 & 0.16/0.20/0.22 & 151.8 & $13 \times  8$ \\
$-$23.0 & 22.4 & 22.76 & $-$19.33/$-$19.54/$-$19.66 & 0.18/0.21/0.24 & 178.4 & $11 \times  9$ \\
20.9 & 25.8 & 20.49 & $-$21.59/$-$21.80/$-$21.92 & 1.42/1.72/1.92 & 184.5 & $25 \times 20$ \\
$-$21.5 & $-$25.4 & 22.71 & $-$19.38/$-$19.59/$-$19.70 & 0.18/0.22/0.25 & 185.0 & $12 \times  8$ \\
\enddata
\tablenotetext{a}{Using Sc/Sa/E $k$-corrections.}
\tablenotetext{b}{Approximate optical extent if at $z=z_{abs}$.}
\tablenotetext{c}{Approximate value due to blending with quasar PSF residuals.}
\tablenotetext{d}{SDSS spectroscopic redshift of $z=0.113$.}
\tablenotetext{e}{Object is only marginally resolved and may be a star.}
\label{table:photometry}
\end{deluxetable}

\begin{deluxetable}{ccccccc}
\tablecaption{Tentative Environment Descriptions}
\tablehead{
\colhead{} & \colhead{} & \colhead{} & \colhead{Galaxy Overlaps} & \colhead{Evidence} & \colhead{$L \ge 4L^*$ and} & \colhead{}\\
\colhead{Sightline} & \colhead{$W_0^{\lambda2796}$} & \colhead{$z_{abs}$} & \colhead{Sightline?} & \colhead{for Pair?} & \colhead{$b<90$ kpc?\tablenotemark{a}} & \colhead{Category}\\
}
\startdata
0013+1414   & 2.69 & 0.484 & no    & yes   & yes & norm./inter. \\
0240$-$0812 & 2.91 & 0.531 & no    & no    & no  & normal \\
1038+4727   & 3.14 & 0.529 & no    & close & no  & normal \\
0836+5132   & 3.48 & 0.567 & close & no    & no  & normal \\
0747+3054   & 3.63 & 0.765 & no    & yes   & yes & bright/inter. \\
0800+2150   & 3.65 & 0.572 & no    & yes   & no  & norm./inter. \\
0232$-$0811 & 3.69 & 0.452 & yes   & no    & no  & overlap \\
0902+3722   & 3.97 & 0.670 & no    & yes   & no  & norm./inter. \\
1520+6105   & 4.24 & 0.423 & close & no    & no  & normal \\
1427+5325   & 4.35 & 0.554 & yes   & yes   & no  & overlap/inter. \\
0747+3354   & 4.69 & 0.620 & yes   & close & no  & overlap \\
1011+4451   & 4.94 & 0.836 & no    & close & yes & bright \\
1000+4438   & 5.33 & 0.719 & yes   & no    & no  & overlap \\
1417+0115   & 5.6  & 0.669 & no    & no    & yes & bright \\
1356+6119   & 5.97 & 0.785 & yes   & no    & no  & overlap \\
\enddata
\tablenotetext{a}{For $z_{gal}=z_{abs}$}
\label{table:environment}
\end{deluxetable}

\clearpage

\clearpage


\clearpage

\end{document}